\documentclass[fleqn,usenatbib]{mnras}
\usepackage{newtxtext,newtxmath}
\usepackage[T1]{fontenc}
\usepackage{ae,aecompl}

\usepackage{graphicx}	% Including figure files
\usepackage{amsmath}	% Advanced maths commands
\usepackage{amssymb}	% Extra maths symbols
\usepackage{ulem}
\usepackage{xcolor}
\usepackage{arydshln}

\title[Baryon acoustic oscillation measurement errors]{How accurately can we measure the baryon acoustic oscillation feature?}

\author[Ruggeri \& Blake]{
\parbox{\textwidth}{
Rossana Ruggeri$^{1}$\thanks{Email: rruggeri@swin.edu.au}, Chris Blake$^{1}$
}
\vspace*{15pt} \\
$^{1}$ Centre for Astrophysics $\&$ Supercomputing, Swinburne University of Technology, P.O. Box 218, Hawthorn, VIC 3122, Australia \\
}
\date{Accepted XXX. Received YYY; in original form ZZZ}

\pubyear{2019}

\begin{document}

%\label{firstpage}
\maketitle

\begin{abstract}
Baryon acoustic oscillations (BAO) represent one of the cleanest probes of dark energy, allowing for tests of the cosmological model through the measurement of distance and expansion rate from a 3D galaxy distribution.  The signal appears at large scales in the correlation function where linear theory applies, allowing for the construction of accurate models.  However, due to the lower number of modes available at these scales, sample variance has a significant impact on the signal, and may sharpen or widen the underlying peak.  Therefore, equivalent mock realizations of a galaxy survey present different errors in the position of the peak when uncertainties are estimated from the posterior probability distribution corresponding to the individual mocks.  Hence the posterior width, often quoted as the error in BAO survey measurements, is subject to sample noise.  A different definition of the error is provided by the asymptotic variance of the maximum likelihood estimator, which involves the average over multiple realizations, and is not subject to sample noise.  In this work we re-analyse the main galaxy survey data available for BAO measurements and quantify the impact of the noise component on the error quoted for BAO measurements.  We quantify the difference between three definitions of the error: the confidence region computed from a single posterior, the average of the variances of many realizations, and the Fisher matrix prediction assuming a Gaussian likelihood.
\end{abstract}


%%%%%%%%%%%%%%%%%%%%%%%%%%%%%%%%%%%%%%%%%%%%%%%%%%

%%%%%%%%%%%%%%%%% BODY OF PAPER %%%%%%%%%%%%%%%%%%
\section{Introduction}

In the early pre-recombination Universe, photon and baryon coupling gives rise to acoustic oscillations, as the infall of baryons into dark matter potential wells is balanced by radiation pressure.  At decoupling, photons diffuse away and, shortly after recombination, the baryons are left distributed in shells of various sizes whose characteristic scale is defined by the physical size of the sound horizon $r_d$,  
\begin{eqnarray}
r_d = \int_{z_d}^{\infty} \frac{c_s(z)}{H(z)}\mathrm{d}z, 
\end{eqnarray}
where $c_s$ is the sound speed, $H(z)$ is the Hubble parameter, and $z_d$ is the redshift of the baryon drag epoch.  A detailed description of the physics is available in \cite{2013PhR...530...87W}.  

Observations of the large-scale structure of the galaxy distribution reveal the imprint caused by these pressure waves in late-time clustering. The imprint is evident as a localized peak in the correlation function at the characteristic scale $r_d$, and as a damped series of oscillations in the power spectrum.  This baryon acoustic oscillation (BAO) signal represents one of the cleanest probes of dark energy.  Assuming standard radiation and matter contributions to the energy density, CMB observations allow the sound horizon to be calibrated within $0.2\%$ \citep{2016A&A...594A..13P}.  The acoustic peak may then be utilised as a standard ruler to measure distances and expansion in the late-time Universe.

Initial measurements of the expansion of the Universe using BAO detections included the Sloan Digital Sky Survey (SDSS) I and II \citep{2005ApJ...633..560E}, the 2-degree Field Galaxy Redshift Survey (2dFGRS) \citep{2005MNRAS.362..505C}, the WiggleZ Dark Energy Survey \citep{2011MNRAS.415.2892B}, and the 6-degree Field Galaxy Survey (6dFGS) \citep{2011MNRAS.416.3017B}.  More recently, the Baryon Oscillation Spectroscopic Survey (BOSS) obtained the first BAO measurement at percent level accuracy \citep{2014MNRAS.441...24A,2015ApJS..219...12A}.  Furthermore, the Lyman-alpha forest measured in BOSS quasars \citep{2013JCAP...04..026S} and the extended Baryon Oscillation Spectroscopic Survey (eBOSS) \citep{2018MNRAS.473.4773A} have detected the BAO peak at higher redshifts $z > 1$. 

Measurements of cosmological distances from these surveys, combined with Planck measurements of the benchmark model for our Universe, posits the matter content of our Universe to be $\Omega_m \sim 0.31$ and favours a cosmological constant $\Lambda$ for describing the apparent acceleration in the cosmic expansion \citep{2016A&A...594A..13P}.  Larger data sets provided by future experiments such as the Dark Energy Spectroscopic Instrument (DESI) \citep{2019arXiv190710688L} and the Euclid satellite \citep{2011arXiv1110.3193L} aim to improve the constraining power of BAO measurements by a further order of magnitude, due to the unprecedented cosmic volume mapped.

The fact that the BAO peak is imprinted on very large scales $110 \, h^{-1} \mathrm{Mpc}$ allows for an effective linear description of the physics in the model.  However, sample variance has a significant impact on the measurements given the lower number of modes available at these scales, causing different mock realizations of the same galaxy distribution to present different shapes of the BAO peak due to noise fluctuations.  Therefore, if we determine the error in the BAO scale using the posterior probability distribution of each separate realization, we will obtain a different error in the scale.  This variation in the uncertainty across different realizations does not depend on the underlying physics but rather on how sample variance in the data modifies the peak.  If we instead consider the definition of the variance of a maximum-likelihood estimator (MLE), each individual realization should have the same error in the BAO scale, based on an average over all the possible realizations (i.e., Fisher information) which would not be subject to noise.  

The standard BAO fitting approach adopted for existing redshift surveys, e.g.  WiggleZ, 6dFGS, BOSS and eBOSS, considers the posterior probability distribution of the parameterized model fit to the data to determine the maximum-likelihood value and error, but does not quantify the noise component in the results.  In this paper we study the importance of this effect for BAO measurements, and determine the consequences of this noise for published BAO results. In particular we compare three approaches for determining the error in the BAO scale: the variance of the data posterior, the average variance of $N\gg 1 $ mock realizations, and a Fisher matrix estimate. 

This paper is organized as follows: in Sec.\ \ref{sec:2} we review the statistical methods used in this work, with a focus on maximum likelihood estimation.  In Sec.\ \ref{sec:isotropicfit} we review the standard pipeline used to fit the BAO peak in configuration space: in Sec.\ \ref{subsec:corfumod}, \ref{sec:corrfuncmeas} and \ref{sec:error} we review the correlation function template, measurements and errors, and in Sec.\ \ref{subsec:error} we focus on the possible definitions of the error in the BAO scale. In Sec.\ \ref{sec:data} we describe the data samples and mocks used for this study, and present the analysis performed on them. In Sec.\ \ref{sec:fish2} we review the Fisher matrix forecasts for  the errors, and in Sec.\ \ref{sec:discussion} we compare and summarize the results obtained for the different datasets, and the impact of sample noise on previous cosmological results. 

\section{Parameter estimation}\label{sec:2}

In this section we briefly review the general Bayesian approach adopted for data analysis involving parameter estimation.  Extensive discussions of the applications of Bayesian statistics in cosmology are presented by e.g. \cite{2010arXiv1008.4686H} and  \cite{alahheavlect}.

\subsection{Likelihood function}

Given a certain experiment, the goal of parameter estimation is to collocate  the observed data-set  $\mathbf{x}$ within a range of theoretical frameworks based on a number of parameters $\mathbf{\theta}^\mathrm{T} =(\theta_1, \dots \theta_n)$.  From a Bayesian perspective, the goal is to estimate the \textit{posterior} probability, i.e. the probability distribution for each parameter  $\theta$ in the model after setting the priors and running the experiment, $\mathcal{P} (\mathbf{\theta}|\mathbf{x} )$.

Once $\mathcal{P}(\mathbf{\theta}|\mathbf{x})$ is derived, the expectation value and variance of $\theta$ can be computed. However, it is not possible to pre-suppose a form for the posterior distribution, as it depends on the unknown distribution of the data of the experiment.  For this reason it is common to invert the problem, starting with  the expected distribution of the data  given a certain model, named the Likelihood  $\mathcal{L} \equiv  \mathcal{P}(\mathbf{x}|  \mathbf{\theta} )$. If we treat the parameters of the distribution as variables for a fixed  data-set,  we can  assume  a \textit{traditional}\footnote{A multivariate Gaussian function is often adopted.} probability distribution for  $\mathcal{L}$.
 
The relation between $\mathcal{L}$ and the posterior is given by Bayes' theorem,
\begin{equation}\label{eq:bayes}     \mathcal{P} (\mathbf{\theta}|\mathbf{x} ) = \mathcal{L}(\mathbf{x,\mathbf}{\theta} ) \mathcal{P}(\mathbf{\theta})/ \mathcal{P}(\mathbf{x}), 
\end{equation}
where $\mathcal{P}(\mathbf{\theta}) $ represents the prior on the parameters, i.e. our knowledge before the experiment is performed, and $\mathcal{P}(\mathbf{x})$ represents the probability of the data, which enters as a normalization factor for the parameter estimation. Assuming  $\mathcal{P}(\mathbf{x})$ to be a positive constant and the prior on each $\theta_i$ to be flat, it follows from Eq. \ref{eq:bayes} that, \begin{equation}\label{eq:bayes2}
       \mathcal{P} (\mathbf{\theta}|\mathbf{x} ) \propto \mathcal{L}(\mathbf{x,\mathbf}{\theta} ).
\end{equation}
Hence we can derive the form of the posterior from the form of the Likelihood. 

%The error on each  $\theta_i \in \mathbf{\theta} $ is computed from the $ \mathcal{P}(\mathbf{\theta})$, obtained  marginalizing the likelihood over the entire parameter space, e.g. 
% \begin{equation}
%  \mathcal{P}(\theta_1)   = \int d\theta_2 \dots d\theta_N \mathcal{P}(\theta).
% \end{equation}

\subsection{Markov chain Monte Carlo sampler}\label{sec:mcmc}

The marginal distribution of each parameter $\theta_i$ can be obtained by integrating  the  likelihood, e.g for the $N$th parameter, 
\begin{equation}
  \mathcal{P}(\theta_N|\mathbf{x}) =\int d\theta_1..d\theta_{N-1} \mathcal{P}(\mathbf{\theta}|\mathbf{x}).
\end{equation}
For models depending on a large number of parameters, an efficient way to marginalize the posterior (or likelihood) is provided by the widely-used Markov chain Monte Carlo (MCMC) algorithm, which generates samples in the parameter space which may be used to map the one-dimensional posterior distributions (or likelihood) of each parameter. Further, MCMC does not require the full analytic description of the posterior distribution, but only the probability ratio between pairs of points in the parameter space, as each step uses a proposal distribution.  The algorithm is efficient as it scales linearly with the number of parameters \citep{2018ApJS..236...11H} and can be optimized to account for degeneracies and correlation between the different parameters.  We note that, as discussed in \cite{2018ApJS..236...11H}, MCMC may be used as a sampler of the distributions; for other purposes, such as finding the maximum likelihood point, specific optimizers are required. 

\subsection{Maximum likelihood estimate }\label{sec:fisher}

Once the likelihood function is defined, and the posteriors are sampled using Eq. \ref{eq:bayes} with methods such as the one discussed in Sec. \ref{sec:mcmc}, the mean and variance can be computed from the probability distributions obtained. In parameter estimation another quantity is often quoted: the maximum likelihood estimate (MLE), i.e. the parameters $\hat{\mathbf{\theta}}$ which maximize the likelihood function for a given dataset.  In a Bayesian approach the MLE is interpreted as a particular case of the \textit{maximum a posteriori probability (MAP) estimate} where the priors are flat\footnote{Note that if the priors are not flat or the distributions are non-Gaussian, there may not be correspondence between MAP and MLE.}: from Eq. \ref{eq:bayes2}, it follows that $\hat{\mathbf{\theta}}$ maximises the  $\mathcal{P} (\mathbf{\theta}|\mathbf{x} )$ as well. 

It has been shown that the MLE is asymptotically unbiased, i.e. it converges to the true value of $\theta_i$ for a large dataset with $N\rightarrow \infty$, and that it is also asymptotically the minimum variance estimator \citep{2017arXiv170101467T}.  
Considering the Fisher information matrix $F_{ij}$ for each given data set,
\begin{equation}\label{eq:fisher}
 F_{ij} =  -\left\langle \frac{\partial^2 ln \mathcal{L}}{\partial \theta_i \partial \theta_j} \right\rangle ,
\end{equation}
the variance of the MLE variance is given by,
\begin{equation}\label{eq:fisher2}
    \sigma^2_{\theta_i} = \left( F^{-1} \right)_{ii},
\end{equation}
assuming that the likelihood is described by a multivariate Gaussian in $\theta_1,.. \theta_n$.  If the likelihood distribution is not perfectly Gaussian, it will always hold that,
 \begin{equation}
     \sigma^2_{\theta_i} \geq \left( F^{-1} \right)_{ii},
 \end{equation}
where the equality is reached only in the limit of a large dataset, according to the central limit theorem, i.e. the MLE is asymptotically the best unbiased estimator.
 
Some comments on Eq. \ref{eq:fisher} and Eq. \ref{eq:fisher2}: the Fisher information matrix is computed averaging over the ensemble of statistical realizations, therefore the variance in the MLE is also based on the expectation value of different realizations. 
However, it is common in the literature to infer the error in a measured parameter from the posterior of a single experiment. This fact may be unimportant if analyses consider a large number of independent data points, such that the width of the posterior is insensitive to the noise of a single realization (an example in our context would be a fit to the full shape of the galaxy correlation function).  However, if the width of the posteriors vary significantly between realizations, there may be an \textit{error in the error} inferred from the posterior of an individual realization (an example in our context would be a fit to the position of the BAO peak).
 
\section{ BAO fits }\label{sec:isotropicfit}

\subsection{Correlation function model}
\label{subsec:corfumod}
We now review the standard method for fitting the BAO position in configuration space, using the correlation function $\xi(s)$.  This standard ruler technique fits a template to the correlation function by varying a scale distortion parameter $\alpha$, which measures the shift of the acoustic peak in the data with respect to a fiducial model, and is defined as,
\begin{eqnarray}
\alpha \equiv \frac{ D^{2/3}_{A}(z) \, H^{-1/3}(z) \, r_{d,\mathrm{fid}}}
{ D^{2/3}_{A,\mathrm{fid}}(z) \, H_{\mathrm{fid}}^{-1/3}(z) \, r_{d}} ,
\end{eqnarray}
where $D_A(z)$ is the angular diameter distance, $H(z)$ is the Hubble parameter and the subscript ``$\mathrm{fid}$'' indicates the fiducial model selected for the template \citep{2014MNRAS.441...24A}.

For isotropic BAO analysis, $\alpha$ is  the only cosmological parameter of interest: the remaining  physics is included through free nuisance parameters.  The choice of priors and functional forms used for the nuisance parameters is not trivial: on the one hand, too little freedom in the nuisance parameters may lead to biased constraints on $\alpha$. On the other hand, too many nuisance parameters may reduce the constraining power of the data, cancelling the  signal of the acoustic peak.  Following previous analyses (e.g. \cite{2014MNRAS.441...24A}), we test the template with respect to the selection of priors, confirming its robustness.

The adopted model for the correlation function $\xi$ includes non-linear structure formation, scale-dependent galaxy bias and redshift-space distortions in a way that allows us to compute the models efficiently over a range of scales and test a moderate range of cosmologies \citep{2014MNRAS.441...24A}.  The template for $\xi$ is based on the model for the power spectrum at different wave-numbers $k$,
\begin{equation}\label{eq:pmod}
    P (k) = P_{\rm NW} (k)\left[ 1 + \left( \frac{P_{\rm L }(k)}{P_{\rm NW}(k) } -1 \right)e^{-\frac{1}{2} k^2 \Sigma^2_{nl} }\right],
\end{equation}
where $P_{\mathrm NW}$ refers to the no-wiggle power spectrum fitting formulae, derived in \citet{1998ApJ...496..605E}, $P_{\mathrm L}$ refers to the linear-theory power spectrum and the exponential term $ e^{-\frac{1}{2} k^2 \Sigma^2_{nl} } $ takes into account non-linear evolution effects, damping the signal around the BAO scale. 
The  correlation function is obtained by integrating the model in Eq. \ref{eq:pmod} projected on the Bessel function $j_0$, 
\begin{equation}
    \xi (s) = \int \frac{k^2 dk }{2\pi^2} \, P(k) \, j_0(ks) \, e^{-k^2 a^2} ,
\end{equation}
where $a$ is set equal to $1 h^{-1}$Mpc as in previous analyses  \citep{2014MNRAS.441...24A}. 
Four \textit{nuisance} parameters are incorporated in the template, 
\begin{equation}
\xi^{\mathrm {temp}}(s) = B^2 \xi_{\mathrm{mod}} (\alpha s) + A^\xi (s) , 
\end{equation}
where $B$ accounts for the unknown galaxy bias, and, 
\begin{equation}
A(s) = \frac{a_1}{s^2} + \frac{a_2}{s} + a_3 ,
\end{equation}
where $a_1$, $a_2$, $a_3$ are free nuisance parameters, marginalizes over any broadband effects such as redshift-space distortions and non-linear growth of structure.  Following \cite{2014MNRAS.441...24A}, we fix the damping factor $\Sigma_{nl}$ to its best-fit value (obtained on the mocks) for each survey considered.

\subsection{ Correlation function measurement}\label{sec:corrfuncmeas}

We measure the correlation function of each galaxy survey using an efficient algorithm (R. Ruggeri, in prep.) based on the Landy $\&$ Szalay estimator \citep{1993ApJ...412...64L}. Briefly, the estimator considers the galaxy distribution to be a discrete   sample of a continuous underlying  distribution selected by Poisson probability. The correlation at a particular scale $s$ is computed by counting the number of galaxy and randomly-distributed pairs with separation $s$, $DD(s)$ and $RR(s)$, in the data catalogue and a synthetic catalogue built using the  selection function of the data sample with objects randomly located.  The random catalogue is generated with a higher number density (20-100$\times$) than the data catalogue, to limit the Poisson noise component. The pair counting is performed on data $DD$, random $RR$ and cross-data-random $DR$ catalogues to compute the correlation function for each scale $s$ as,
\begin{equation}
    \xi(s) = \frac{DD(s) -2DR(s) + RR(s)  }{RR(s)} .
\end{equation}

\subsection{Clustering errors}\label{sec:error}

Different sources of error contribute to measurements of the correlation function $\xi$, or its Fourier-space counterpart the power spectrum $P$, from a galaxy distribution.  The statistical error can be classified into two components: first, the ``sample (cosmic) variance'', which characterizes the variation in cosmic structure in different regions of the Universe.  In a Fourier-space picture, this structure can be expressed in terms of a sum over underlying modes, where in Gaussian statistics the intrinsic uncertainty on each mode is proportional to $P(k)$ and the number of available modes scales as the volume of the survey.  The second component arises due to the discreteness of the galaxy field (assumed to be Poissonian) which contributes $1/N$ to the total error in each mode, where $N$ is the number of galaxies observed.  This ``shot-noise'' error depends on the observed sample and becomes relatively more important at small scales.  The sample variance component is relatively more important at large scales, and can be reduced by increasing the effective volume of the survey, in order to increase the number of large-scale modes. 
 
We also note that different systematic errors may degrade the measured signal: related to the choice of the estimator, or errors in the calibration of the  instrumentation, imaging pipeline or sky condition during the observations.
The measurements are also affected by dynamical biases such as peculiar velocities contaminating the redshift-space position of galaxies, or the galaxy bias which relates the galaxy field to the matter distribution. 

\subsection{ Defining the error in the BAO scale }\label{subsec:error}

BAO analyses typically determine the error in the fitted scale from the marginalized posterior probability distribution of the scale distortion parameter $\alpha$ fit to the data.  However, as discussed in Sec. \ref{sec:2}, the variance of the MLE  is defined asymptotically as the average of $N$ realizations for $N\rightarrow \infty$.  In contrast with other probes such as growth rate measurements, which are determined from the overall shape of a function, the detection of the BAO feature can be significantly affected by sample noise in each realization.  In this sense, different statistical realizations of a dataset with equivalent effective volume can produce detections of the BAO peak with a range of different significances.  

We can specify several possible descriptions of the error in the scale distortion parameter $\alpha$:

\begin{itemize}
    \item \textbf{Case 1.} \textit{The width of the 68\% confidence region of the sampled posterior probability of the parameter fit to the dataset.}
Note that this estimator of the error does not take into account that the chi-squared distribution of each individual mock/data sample depends on the random noise of the realization.  In order to compute the variance of the MLE with negligible noise, one should consider the asymptotic average over all possible realizations of the measurement.

\item \textbf{Case 2.} Following from Eq. \ref{eq:fisher} and Eq. \ref{eq:fisher2}, the \textit{average over all possible realizations} can be approximated by considering a large number of mock realizations.  We consider estimating the global error using the realizations in two ways:
\begin{itemize}
    \item \textbf{Case 2a.} As the standard deviation of the maximum-likelihood $\alpha$ values fitted to each realization, 
    \item \textbf{Case 2b.} As the average of the width of the 68\% confidence regions of the posterior probability distributions of each realization. 
\end{itemize}
\item \textbf{Case 3.} Finally, we can estimate the error using the Fisher matrix, which  relies on the Gaussian propagation of the error. From Eq. \ref{eq:fisher2},  
\begin{equation}\label{eq:fishalpha}
    \sigma^2_{\alpha} = \left( F^{-1} \right)_{\alpha\alpha}.
\end{equation} 
\end{itemize}
As we discussed in Sec.\ref{sec:2}, the error in Equation \ref{eq:fishalpha} is the lowest possible error achievable using the MLE in the ideal case that the data is Gaussian-distributed.  The Fisher matrix forecast is widely used to predict errors before the data has been collected or if N-body mocks are unavailable \citep{2007ApJ...665...14S, 2014JCAP...05..023F}.

\section{Galaxy Survey Analysis}\label{sec:data}

In this section we summarize the data, mocks and analysis choices for each galaxy survey we consider.  We note that we do not apply density-field reconstruction to any of the datasets.

\subsection{WiggleZ Dark Energy Survey}

\subsubsection{Data and mocks}

The WiggleZ Dark Energy Survey \citep{2010MNRAS.401.1429D}, completed in 2011, collected the redshifts of over $200{,}000$ emission-line galaxies in the southern sky.  The survey was split into 6 regions (9hr; 11hr; 15hr;  22hr; 1hr; 3hr) and the data were analysed in 3 overlapping redshift slices (0.2-0.6, 0.4-0.8, 0.6-1.0).  BAO distance measurements determined from the WiggleZ Survey data were presented by \cite{2011MNRAS.415.2892B} and \cite{2014MNRAS.441.3524K}.

Our chosen fiducial cosmology is different from the one  selected by  \cite{2014MNRAS.441.3524K}: we choose a flat $\Lambda$CDM model with matter density $\Omega_m = 0.31$, baryon density $\Omega_b =0.0448$, Hubble parameter $h =0.71$;  rms clustering amplitude over spheres of $8 h^{-1}$Mpc  $\sigma_8 =0.8$ and  spectral index $n_s = 0.963$. The analysis measures the correlation function at three effective redshifts (0.44, 0.6, 0.73) for a full sample of 158,741 bright emission-line galaxies, after cuts for contiguity.

We also used the 600 WiggleZ Survey mocks described by \cite{2016MNRAS.459.2118K}, which have a  fiducial cosmology slightly different from the one defined above: $\Omega_m = 0.27$, $\Omega_b h^2 = 0.0226 $ with $h = 0.71$.  These 600 mock catalogues were produced using the Lagrangian Comoving Acceleration (COLA) method \citep{2013JCAP...06..036T}.  Before the application of the selection function, the mocks were generated in a $600 h^{-1}$ Mpc cube using $1296^3$ particles of mass $7.5 \times 10^9 h^{-1} M_\odot$.  The mocks were then populated with galaxies using a central and satellite halo occupation distribution tuned to match the number density and clustering of the sample.  Full details are provided by \cite{2016MNRAS.459.2118K}.

\subsubsection{Analysis}
\label{subsec:wigglefit}
 \begin{figure}
     \centering
     \includegraphics[width = \columnwidth]{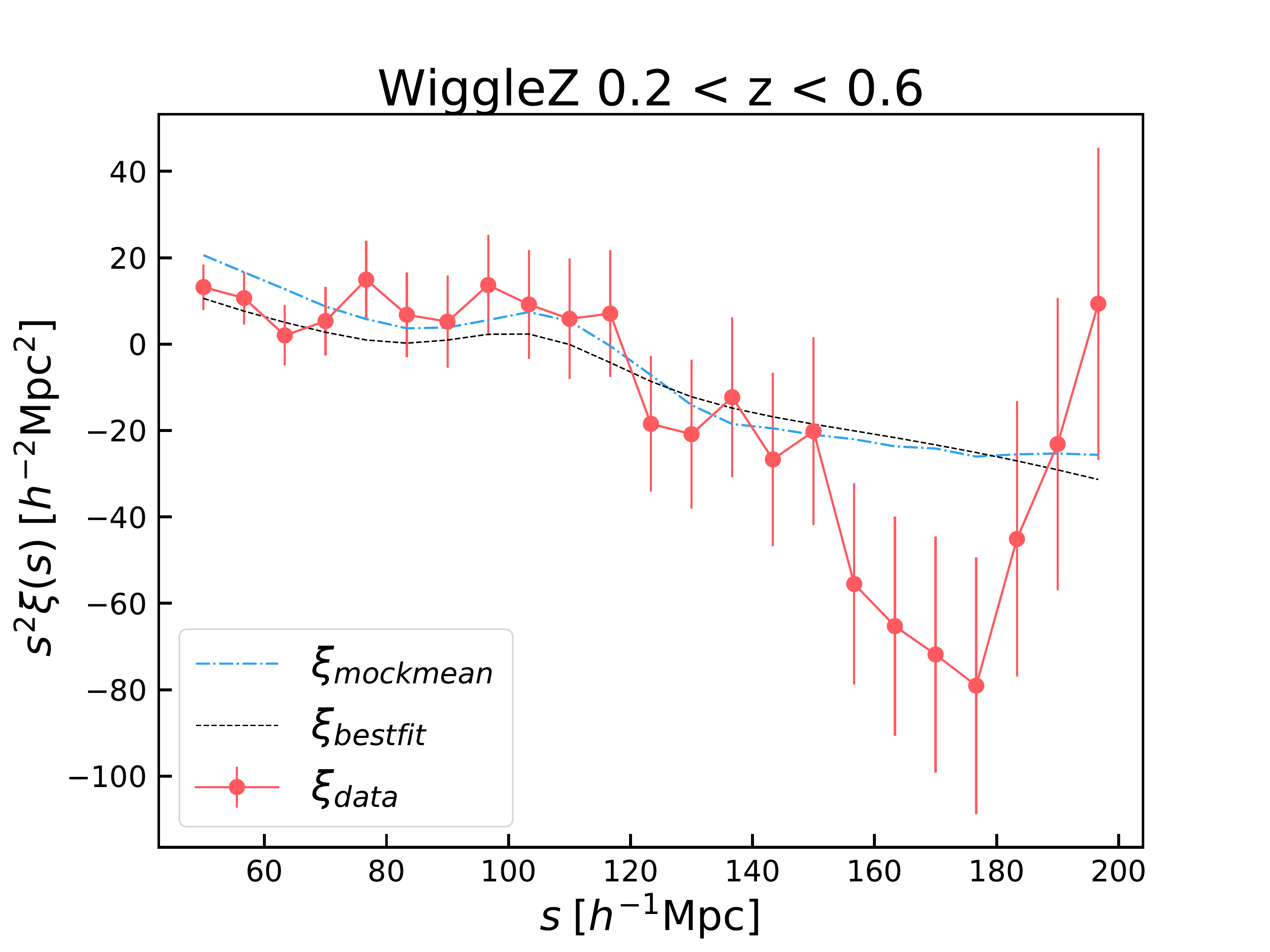}
     \includegraphics[width = \columnwidth]{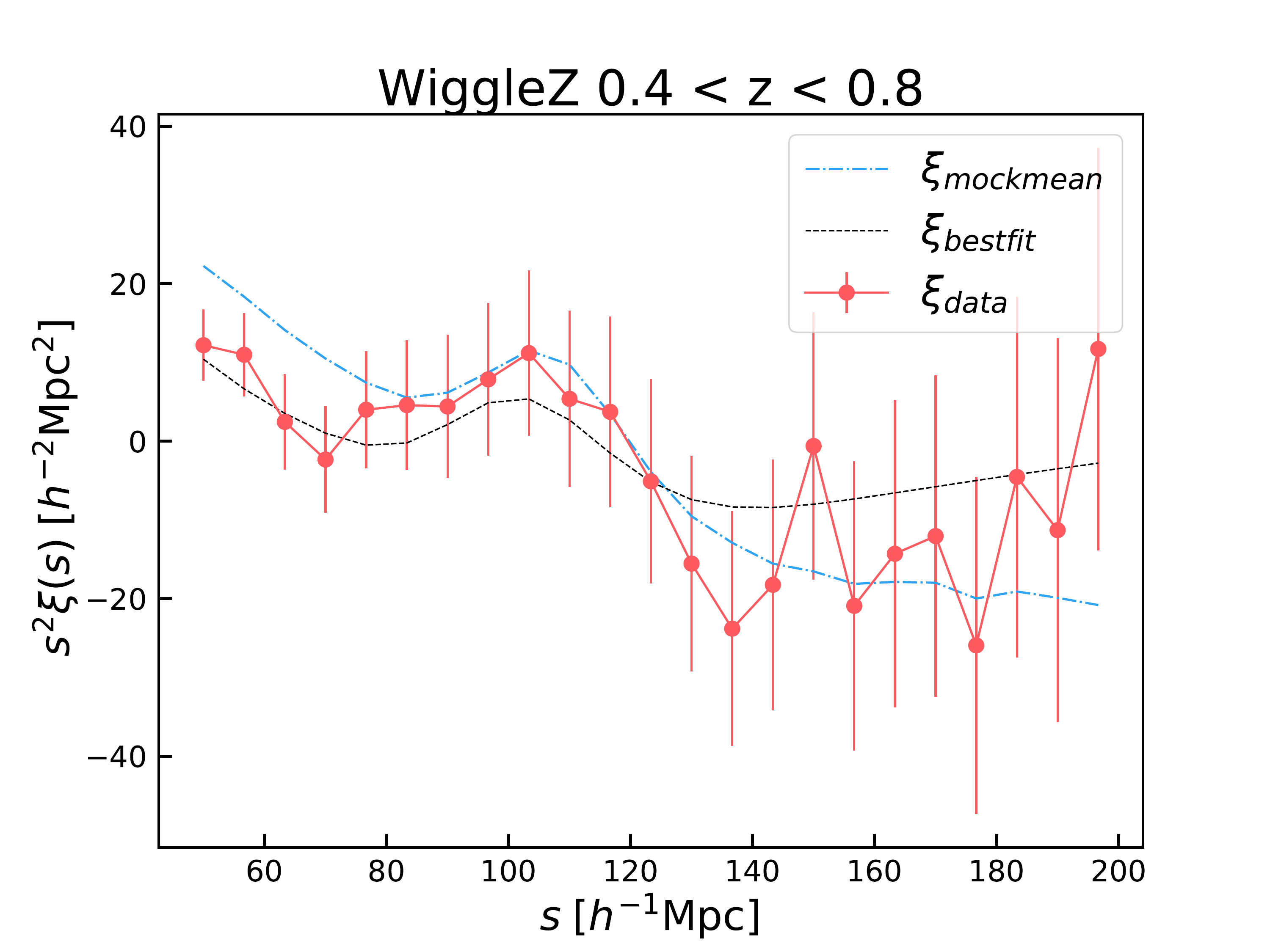}
     \includegraphics[width = \columnwidth]{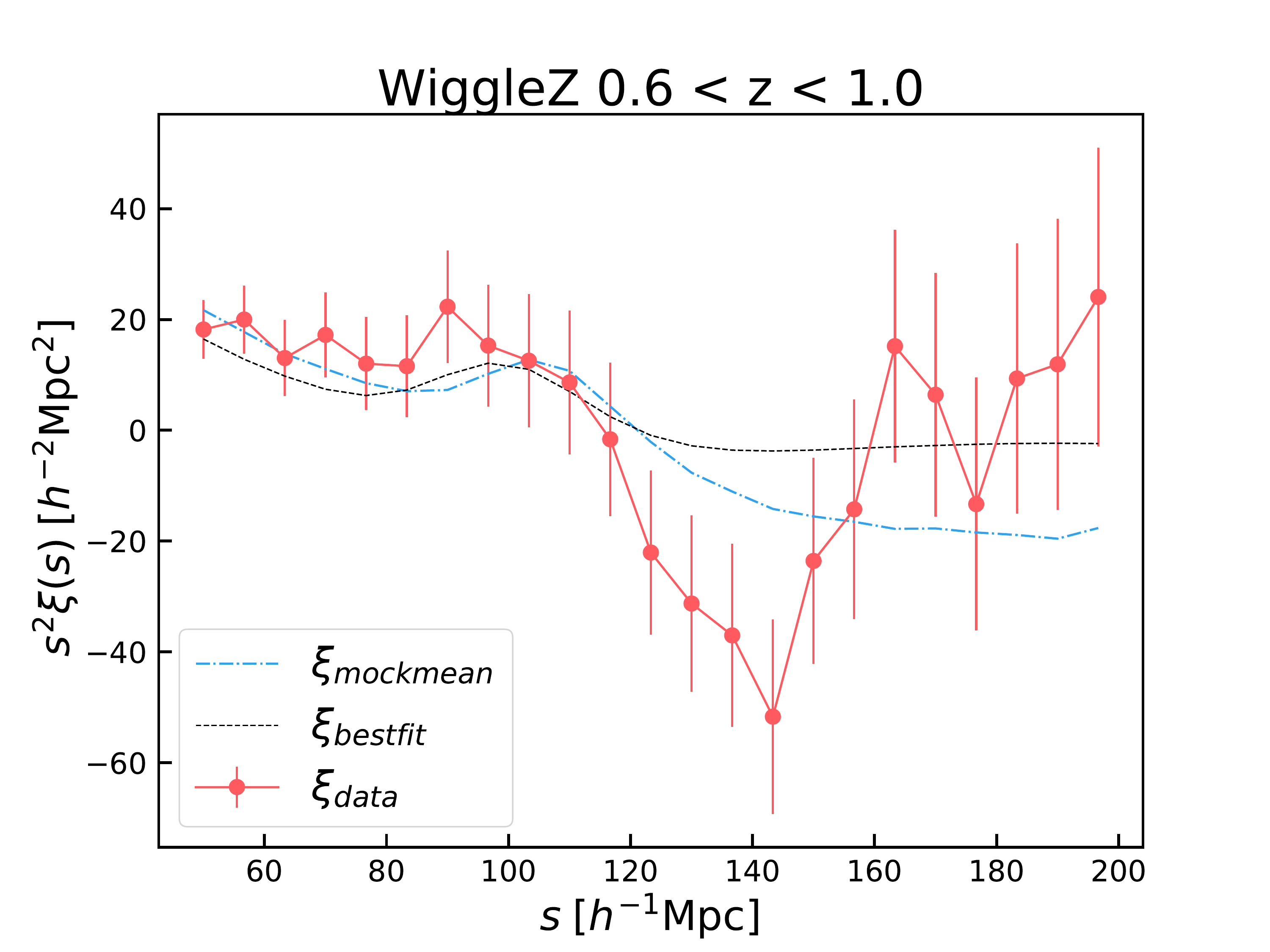}
          \caption{The correlation function of the WiggleZ Dark Energy Survey in three redshift ranges (0.2-0.6, 0.4-0.8, 0.6-1.0) (top, middle, bottom panels).  Red solid line: the correlation function of the data for scales between $40-200 \, h^{-1}$ Mpc. For each redshift slice the correlation function is given as the weighted average of correlation functions measured in 6 different regions.  The blue dash-dot lines correspond to the mock mean correlation functions.  The black dashed lines correspond to the best-fitting model for $\xi$. %\cab{[Suggest changing ``mean'' in legends to ``mockmean'' for clarity]}
          }
     \label{fig:wiggle1}
 \end{figure}

Following the methodology presented by \citet{2014MNRAS.441.3524K}, for each redshift slice ($0.2-0.6$, $0.4-0.8$, $0.6-1.0$) we measured the correlation function for the data and each mock for each region ($m=1\dots 6$), using the estimator described in Sec. \ref{sec:corrfuncmeas}.  Following \cite{1994ApJ...426...23F}, we applied ``FKP weights'' to each galaxy to enhance the signal-to-noise, $w_{\mathrm{FKP}} = 1/[1 + n(z)P_0]$, where $n$ is the number density and $P_0 = 5000 h^{-3}$ Mpc$^3$ is the power spectrum amplitude on the scales of interest.

We measured the correlation function in separation bins between 40 and 200 $h^{-1}$ Mpc of width 6.7 $h^{-1}$ Mpc.  Figure \ref{fig:wiggle1} displays the measured correlation functions from the data (red solid lines) and mock mean (blue dash-dotted line) in the $0.2 <z<0.6$, $0.4<z<0.8$ and $0.6<z<1.0$ redshift slices.

We derived the covariance matrix of the correlation function measurement in each region $m$ from the mocks as,
\begin{equation}\begin{split}
C_{m, (i,j)} = \frac{1}{N -1 } \sum_{n=1}^{N} \big[ \xi_{m, n }(s_i) - \hat{\xi}_m(s_i) \big]  \times \big[  \xi_{m, n}(s_j) - \hat{\xi}_m(s_j) \big],
\label{covmat}
\end{split}\end{equation}
where $\hat{\xi}$ is the mean correlation function of the $N = 600$ mock realizations. We then optimally combined the measurements in the 6 regions following the procedure described by \citet{2011ApJ...728..126W}, where the combined correlation function is given by,
\begin{equation}
\xi = C \sum_m (C_m)^{-1}\xi_m ,
\end{equation}
and the combined covariance $C$ is related to the individual region covariances as,  
  \begin{equation}
  C^{-1} = \sum_m C^{-1   }_m.
  \end{equation}
We applied the fitting pipeline described in Sec \ref{sec:isotropicfit} to the combined correlation function in each redshift slice using 5 free parameters: $\alpha$ and the nuisance parameters $a_1$, $a_2$, $a_3$, $B$.
We performed the fits using an efficient MCMC sampler for the likelihood, assumed to be Gaussian.  The following priors are defined for $\alpha$ and $B$: $0.8 < \alpha < 1.2 $, $0.0 < B$. $$ \;$$ $$\;$$ $$\;$$ $$\;$$

The convergence  is determined using the \cite{Gelman:1992zz} criteria based on independent chains. $\Sigma_{NL}$ is fixed to the best-fitting value to the mean of the mocks for each redshift ranges considered  $\Sigma_{NL} =9.6, 9.2, 9.8  h^{-1}$ Mpc, while we place wide, flat priors on the other parameters.  We summarize the results obtained for the three redshift ranges in Table \ref{tab:resultALL}, where $\alpha$ and $\sigma_\alpha$ are determined using the mean and width of the posterior probability distributions of the data. The reduced chi-square statistic $\chi^2/D.O.F.$ is also reported: the number of degrees of freedom, $D.O.F.$, is computed by subtracting the number of free parameters (5) from the number of separation bins (23).

\begin{table}
%\begin{minipage}[t]{0.48\linewidth}  
\begin{tabular}{ccccc}
   \hline
   Survey & Redshift range & $\alpha$ & $\sigma_\alpha$  &  $\chi^2/D.O.F.$ \\
   \hline
  WiggleZ &  $0.2< z < 0.6$   &1.02 & 0.10  & 25/(23-5)\\
  WiggleZ &  $0.4< z < 0.8$   &1.04 & 0.078 & 15/(23-5)\\
  WiggleZ &  $0.6< z < 1.0$   &1.07 & 0.074 & 20/(23-5)\\
   & & & & \\
  6dFGS    &  $0 < z < 0.2$   &1.06  & 0.072  & 10/(17-5)\\
   & & & & \\
 BOSS   &  $0.43< z < 0.7$   &0.995 & 0.015  & 18/(24-5)\\
   & & & & \\
  eBOSS  &  $0.8 < z < 2.2$   &0.98  & 0.086  & 15/(22-5)\\
   \hline
    \end{tabular}
  \caption{Results of BAO fits to the WiggleZ, 6dFGS, BOSS and eBOSS surveys. The $\alpha$ and $\sigma_\alpha$ values reported correspond to the mean and half-width of the 68$\%$ confidence regions of the data posteriors. The degrees of freedom $D.O.F.$ associated with the chi-squared statistic $\chi^2$ correspond to the number of separation bins minus the $5$ free parameters in the template.
  %\cab{[I suggest adding $D_V/r_d$ values and errors to this table]}
  }
    \label{tab:resultALL}
 \end{table}

\begin{figure}
    \centering
    \includegraphics[width= \columnwidth]{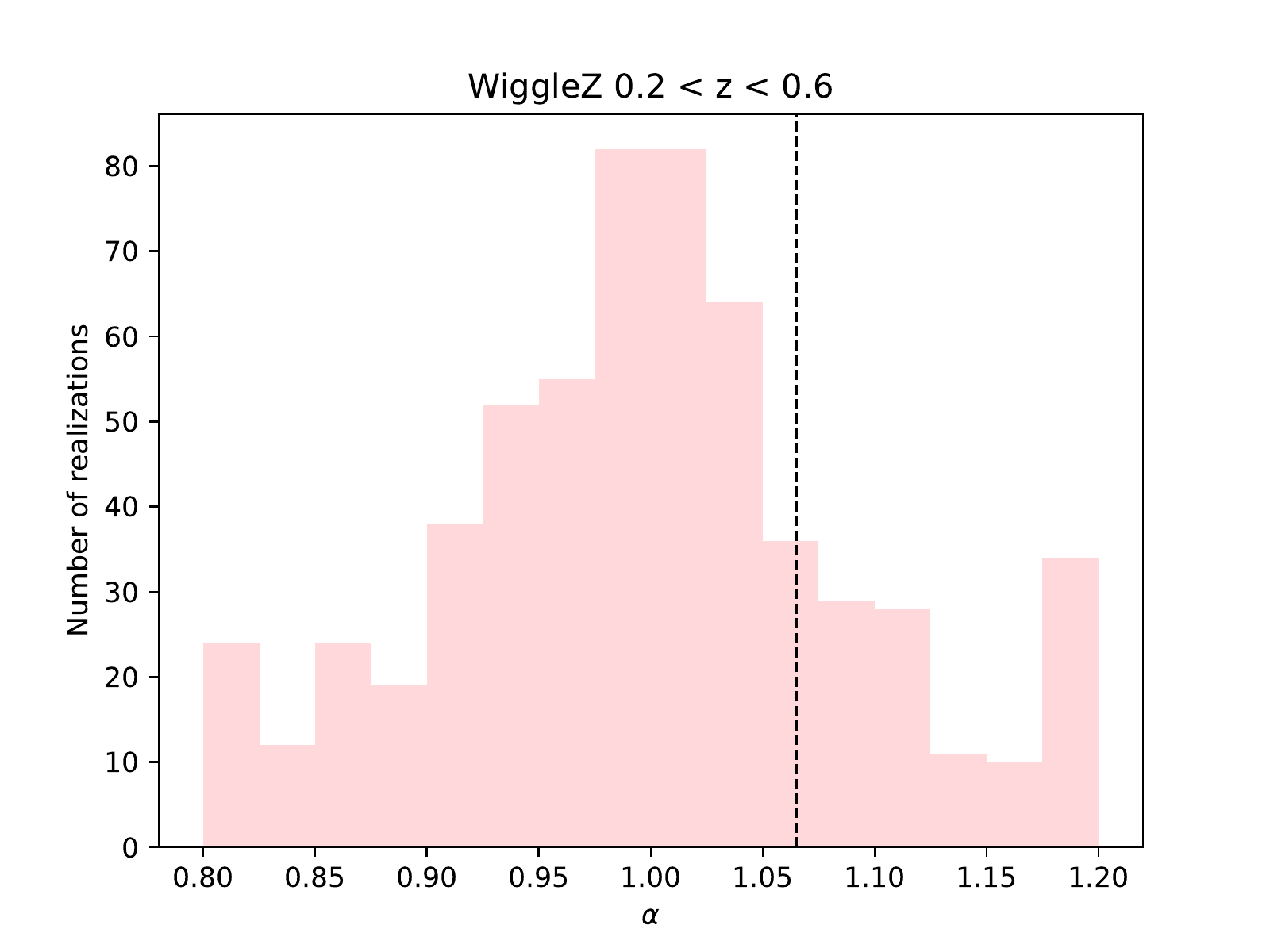}
        \includegraphics[width= \columnwidth]{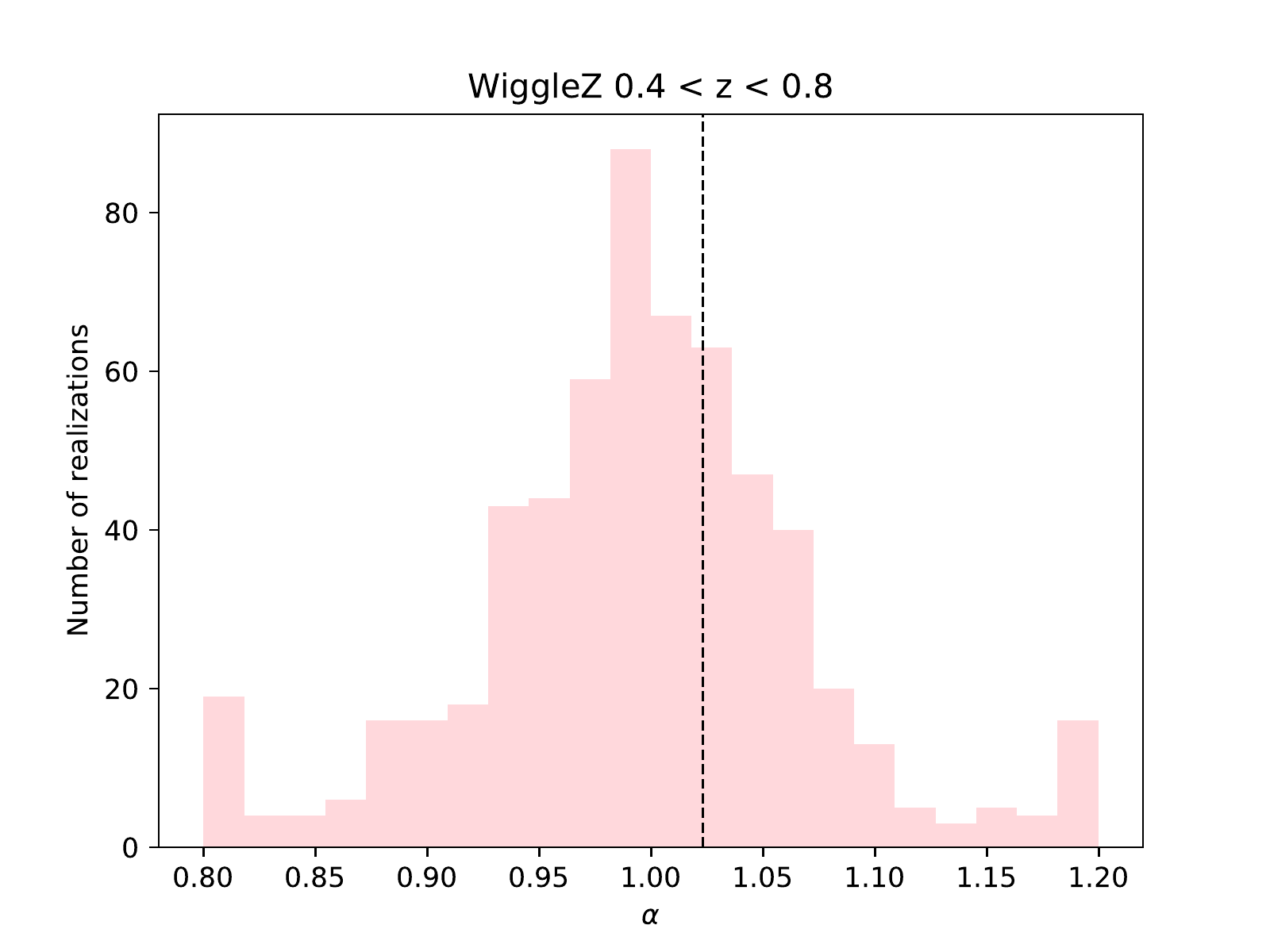}
        \includegraphics[width= \columnwidth]{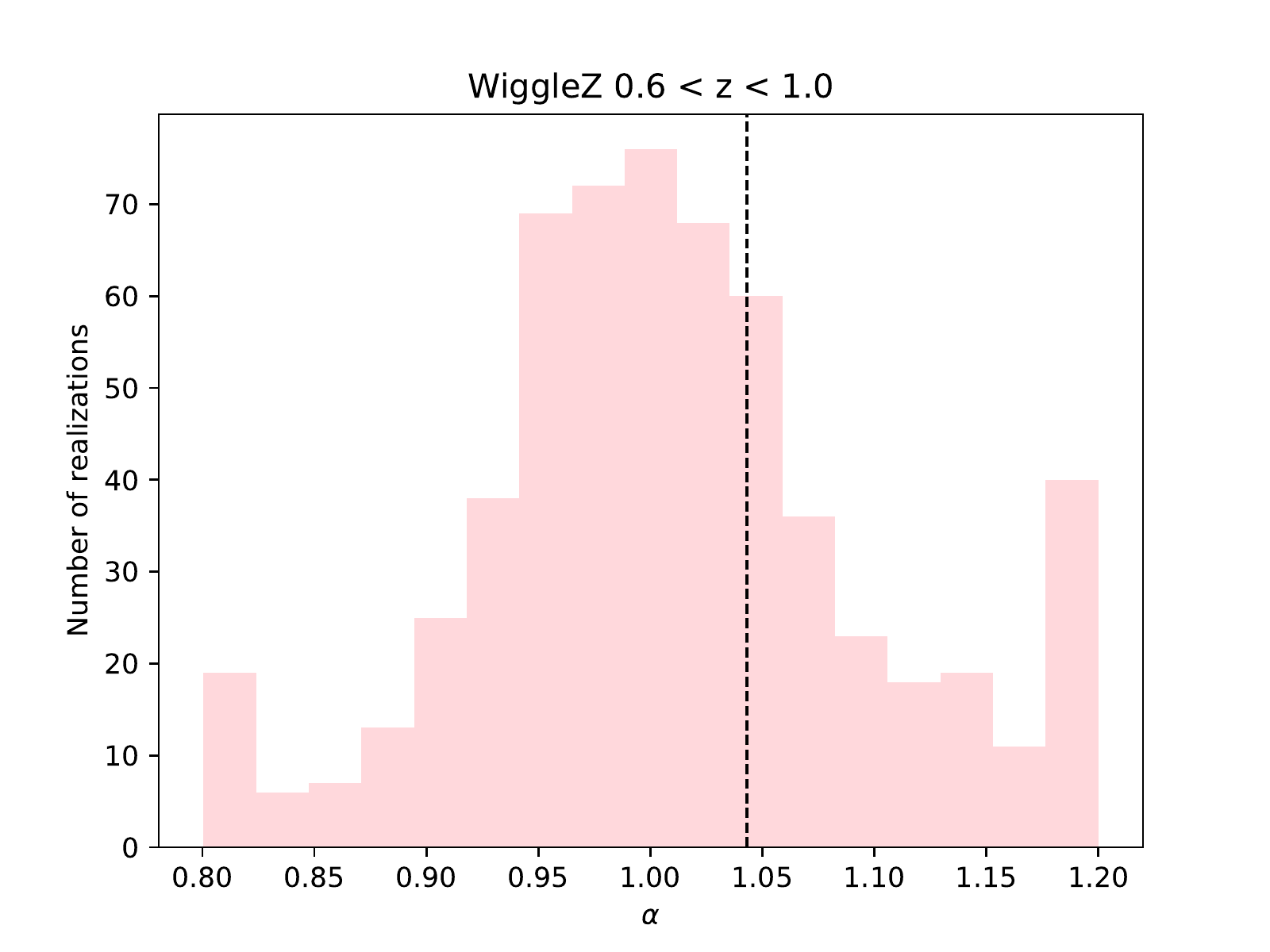}
    \caption{Best-fitting BAO distortion parameters $\alpha$ for the WiggleZ Dark Energy Survey  mocks for redshift bins $0.2 < z < 0.6$, $0.4 < z < 0.8$ and $0.6 < z < 1.0$ (top, middle, bottom panels respectively). Each $\alpha$ is computed as the mean value of the posterior obtained from each mock.  The dashed black line displays the  mean $\alpha$ value obtained from the data posterior in each case. 
  %  \cab{[Suggest re-formatting this figure with $x-$ and $y-$axis labels like Fig.5 and panel titles like Fig.1]}
    }
    \label{fig:alphawiggle}
\end{figure}

\begin{figure}
    \centering
    \includegraphics[width= \columnwidth]{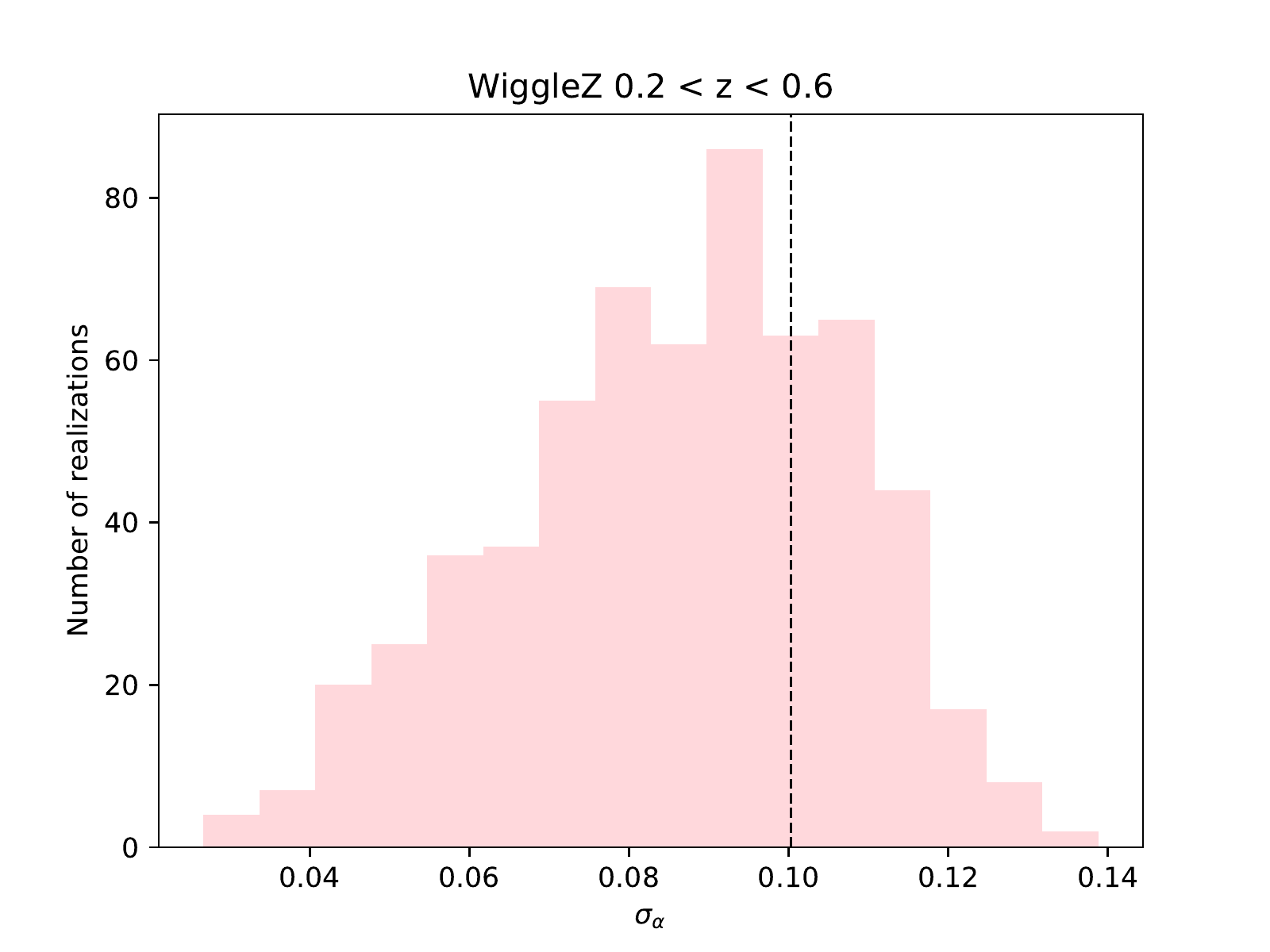}
        \includegraphics[width= \columnwidth]{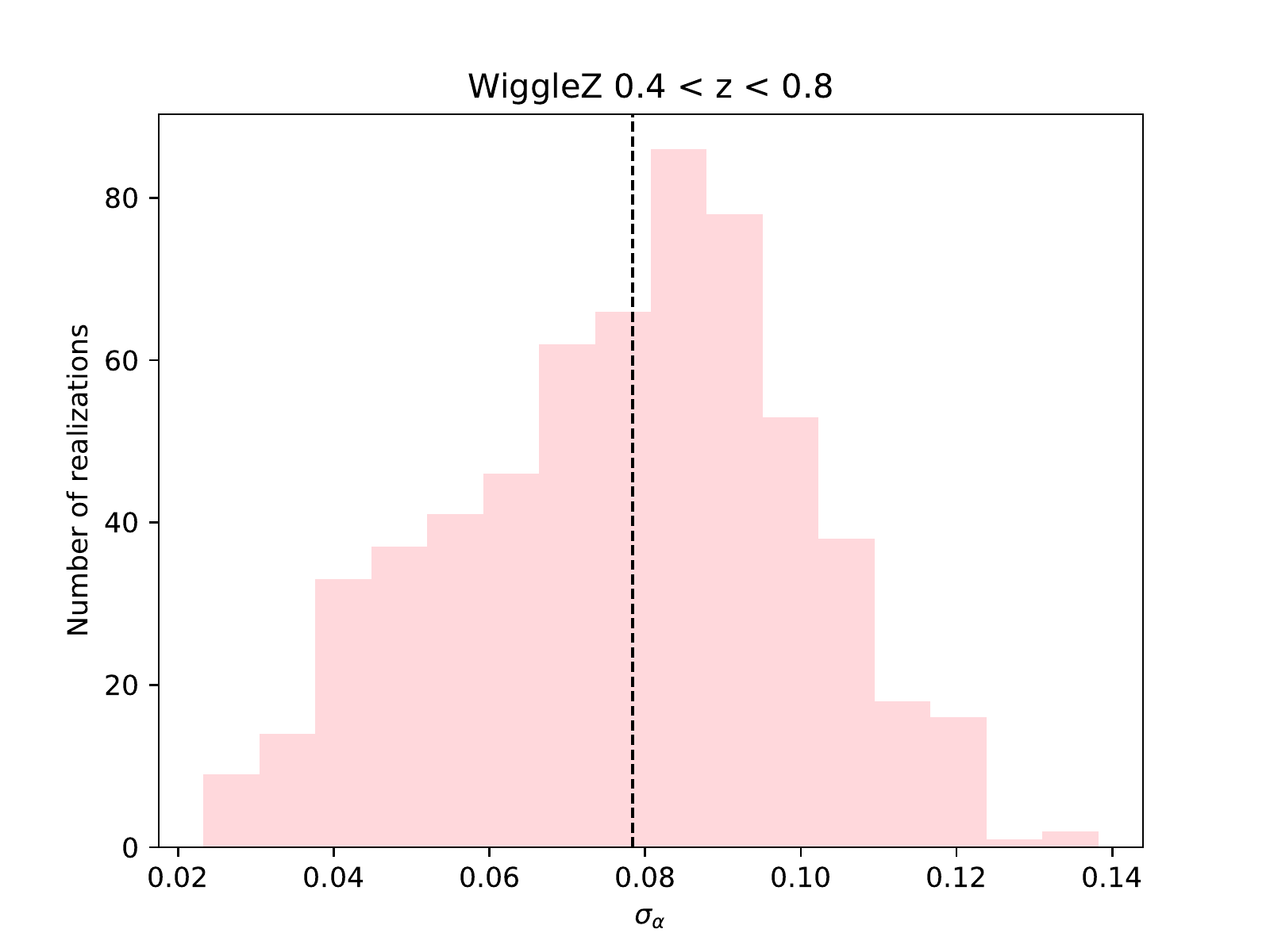}
        \includegraphics[width= \columnwidth]{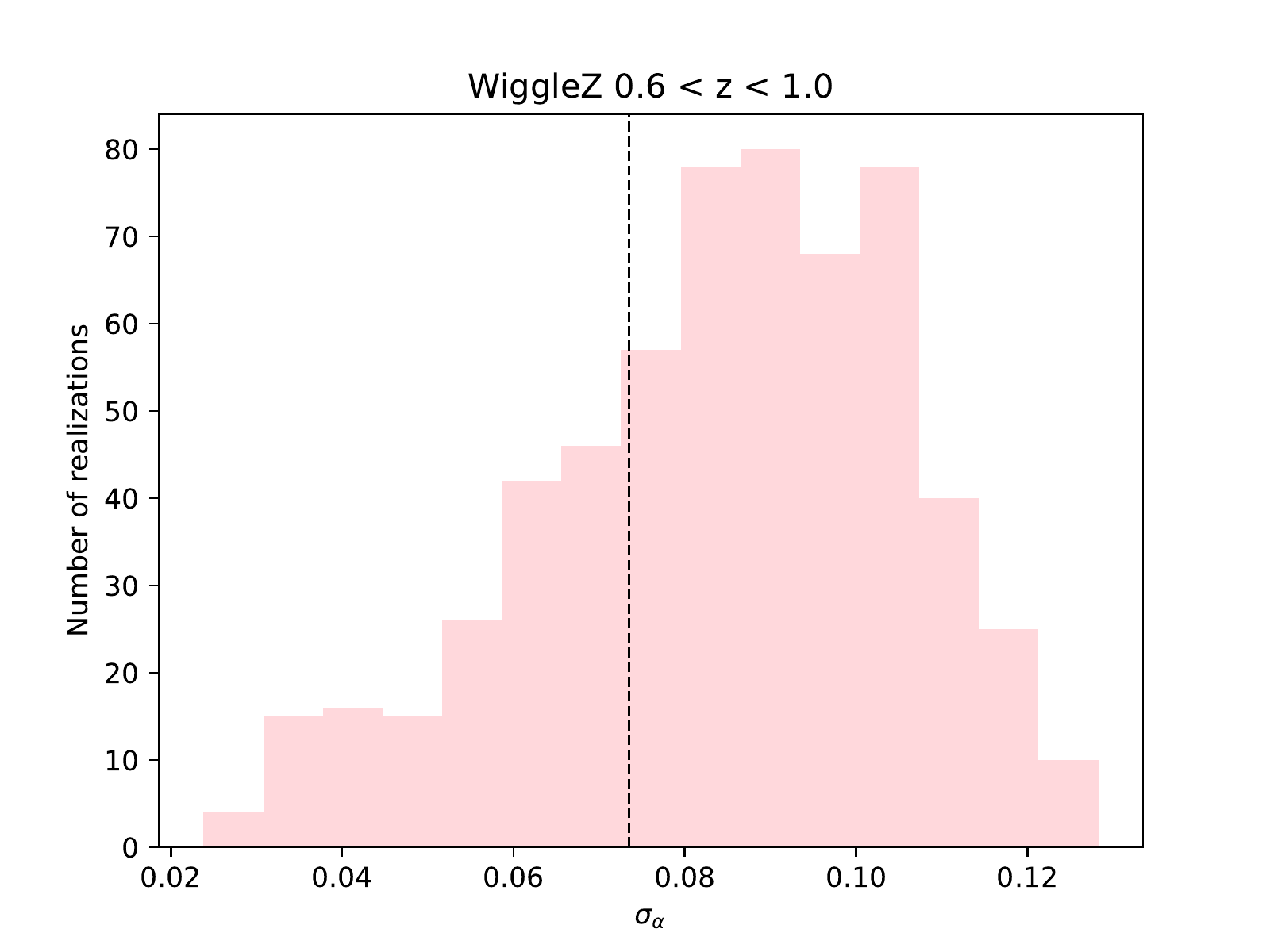}
  \caption{Width of the posterior probability distributions for $\alpha$ for the WiggleZ Dark Energy Survey mocks for redshift bins $0.2 < z < 0.6$, $0.4 < z < 0.8$ and $0.6 < z < 1.0$ (top, middle, bottom panels respectively). Each $\sigma_\alpha$ is evaluated from the width of the $68\%$ confidence region of the resulting posterior probability distribution computed for each mock.  The dashed black line displays the value of $\sigma_\alpha$ obtained using the posterior of the data in each case. %\cab{[Suggest re-formatting this figure with $x-$ and $y-$axis labels like Fig.6 and panel titles like Fig.1]}
  }
    \label{fig:sigalphawiggle}
\end{figure}

Figure \ref{fig:alphawiggle} shows a histogram of the best-fitting values of $\alpha$ for the 600 WiggleZ COLA mocks for the three redshift bins $0.2-0.6$, $0.4-0.8$, $0.6-1.0$ (top, middle, bottom panels).  The dashed vertical line corresponds to the value of $\alpha$ as measured from the data catalogue, which lies within the mock distribution for each redshift range.  The width of the distribution of $\alpha$ in Fig. \ref{fig:alphawiggle} defines the Case 2a error in the BAO fit.  We note that the error estimate in Case 2a corresponds to the average of the posteriors only if individual posteriors are Gaussian; however, due to nuisance parameters and the low signal of the detection, the individual posteriors measured from WiggleZ Survey data and mocks display non-Gaussian features. In Sec \ref{sec:discussion} we discuss this aspect in more detail.
Figure \ref{fig:sigalphawiggle} shows the distribution of the standard deviations, $\sigma_\alpha$, computed from the posterior probability distribution of each mock while  marginalizing over the other parameters $a_1$, $a_2$, $a_3$, $B$.   Each individual measurement of $\sigma_\alpha$ corresponds to a Case 1 error, while the mean of the distribution in Figure \ref{fig:sigalphawiggle} corresponds to the Case 2b error.  The black dotted vertical line indicates the standard deviation of the posterior of the data. Each panel corresponds to one of the three redshift bins $0.2-0.6$, $0.4-0.8$, $0.6-1.0$ (top, middle, bottom). The $0.4-0.8$ bin  corresponds to the sample with the highest S/N, while the $0.6-1.0$ bin contains the lowest number of objects.  The error obtained from the posterior of the data is within the range of errors obtained across the mocks, but does not necessarily match the mean of the distribution, as is particularly evident in the third redshift range considered.  In Sec. \ref{sec:discussion} we quantify these discrepancies for all the error estimates.


\subsection{6-degree Field Galaxy Survey (6dFGS)}

\subsubsection{Data and mocks}

We adopted the same catalogue used by \citet{2011MNRAS.416.3017B} and \citet{2018MNRAS.481.2371C} for their 6dFGS BAO analyses.  This dataset consists of $75{,}117$ galaxies in the nearby Universe with mean redshift $\overline{z} \approx 0.053$ over $17{,}000$ deg$^2$.  The sample is selected using magnitude cut $K < 12.9$ and limiting to sky regions with more than $60\%$ completeness.  Corresponding mock catalogues are created with the COLA method \citep{2013JCAP...06..036T} from N-body simulations with $2.8  \times 10^{10} h^{-1} M_\odot$ mass resolution from a snapshot at $z=0.1$, populating these mocks with a central and satellite halo occupation distribution to match the number density and clustering of the sample.  The fiducial cosmology adopted for the creation of the 6dFGS mocks and the following analysis is a flat $\Lambda$CDM model with parameters $\Omega_m = 0.3$, $\Omega_b = 0.0478$, $h =0.68$, $\sigma_8 = 0.82$ and $n_s = 0.96$.  Random catalogues are produced using the selection function of the data and mocks, selecting a $30\times$ higher number density than the data to minimize the Poisson noise.

\subsubsection{Analysis}\label{subsec:6dfit}
 
\begin{figure}
     \centering
     \includegraphics[width= \columnwidth]{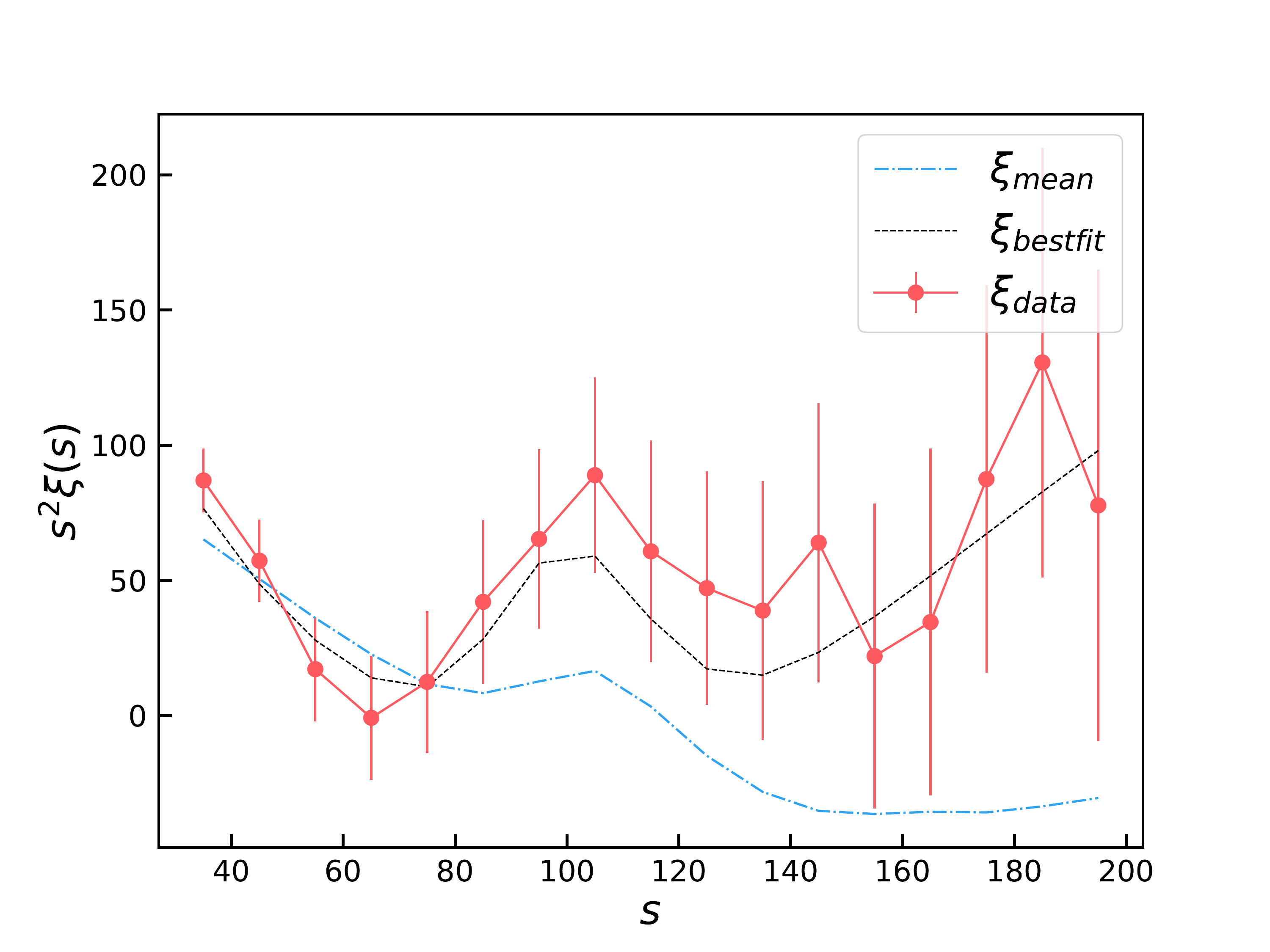}
             \caption{The correlation function of the 6-degree Field Galaxy Survey.  The red solid line shows the 6dFGS data correlation function for scales between $30-200 \, h^{-1}$ Mpc with  $10 \, h^{-1}$ Mpc width bins.  The blue dotted line corresponds to the mean of $600$ COLA mock correlation functions.  The black dashed line is the best-fitting model for $\xi$. %\cab{[Suggest re-formatting this figure with $x-$ and $y-$ axis labels like Fig.1, ``mockmean'' rather than ``mean'' in the legend, and title of panel = ``6dFGS'']}
             }
    \label{fig:6dfcf}
 \end{figure}

We measured the correlation function with the estimator presented in Sec. \ref{sec:corrfuncmeas}, applying FKP weights to individual galaxies with characteristic power spectrum  $P_0 = 10{,}000 \, h^{-3}$ Mpc$^3$.   We used separations in the range $30 < s < 200 \, h^{-1}$ Mpc with $10 \, h^{-1}$ Mpc bin width.  Figure \ref{fig:6dfcf} displays the measured 6dFGS correlation function (red dashed line) compared with the average of the correlation function computed from 600 mocks. From the $600$ correlation functions measured in 17 separation bins, we constructed a $17 \times 17$ covariance matrix,
\begin{equation}\begin{split}
C_{i,j} = \frac{1}{N -1 } \sum_{n=1}^{N} \big[ \xi_{ n }(s_i) - \hat{\xi}(s_i) \big]  \times \big[  \xi_{ n}(s_j) - \hat{\xi}(s_j) \big],
\label{eq:6dfcovmat}
\end{split}\end{equation}
with $N=600$.

We then fit for the BAO feature in the correlation functions for the data and each mock, using the template described in Sec. \ref{sec:isotropicfit}.  The fitting pipeline used the MCMC sampler and convergence criteria described in Sec. \ref{subsec:wigglefit}.  In the same manner as Sec. \ref{subsec:wigglefit}, we fit for $\alpha$ and marginalized over a set of nuisance parameters  $a_1$, $a_2$, $a_3$, $B$, fixing the non-linear damping factor to the value measured from the mean of the mock, $\Sigma_{nl} = 10.3 \, h^{-1}$ Mpc.  We defined wide, flat priors for the nuisance parameters and additionally used the prior $0.7 < \alpha < 1.3$. 

Table \ref{tab:resultALL} summarizes the results of the fit to the 6dFGS data sample; $\alpha$ and $\sigma_\alpha$ are estimated from the mean and width of the data posterior probability distribution.  The reduced chi-squared statistic is also reported with $D.O.F. = 17-5$.

\begin{figure}
    \centering
    \includegraphics[width= \columnwidth]{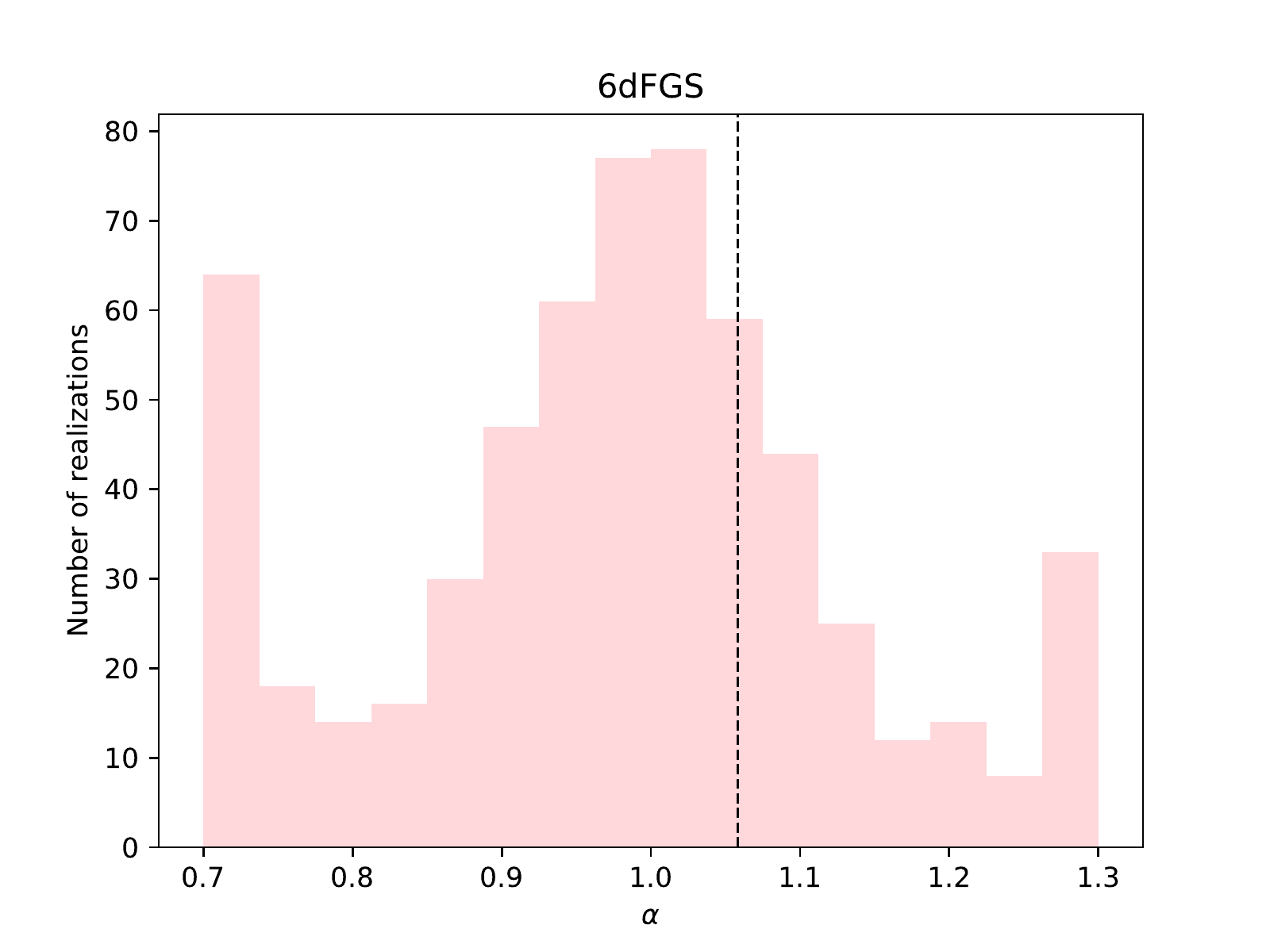}
    \caption{Best-fitting BAO distortion parameters $\alpha$ for the 6dFGS COLA mocks.  Each $\alpha$ is computed as the mean value of the posterior obtained from each mock. The value obtained fitting the data catalogue  corresponds to the dashed vertical line.}
    \label{fig:6dfalpha}
\end{figure}

The histogram in Fig. \ref{fig:6dfalpha} displays the distribution of the MLE values of $\alpha$ for the 600 mocks.  The vertical dashed line denotes the value obtained by fitting the data, in agreement with \cite{2018MNRAS.481.2371C}.  Note that some mock realizations do not yield a significant detection of the BAO peak, hence the build-up of MLE values at $\alpha = 0.7$ and $\alpha = 1.3$, the two boundaries of the prior.

\begin{figure}
    \centering
    \includegraphics[width= \columnwidth]{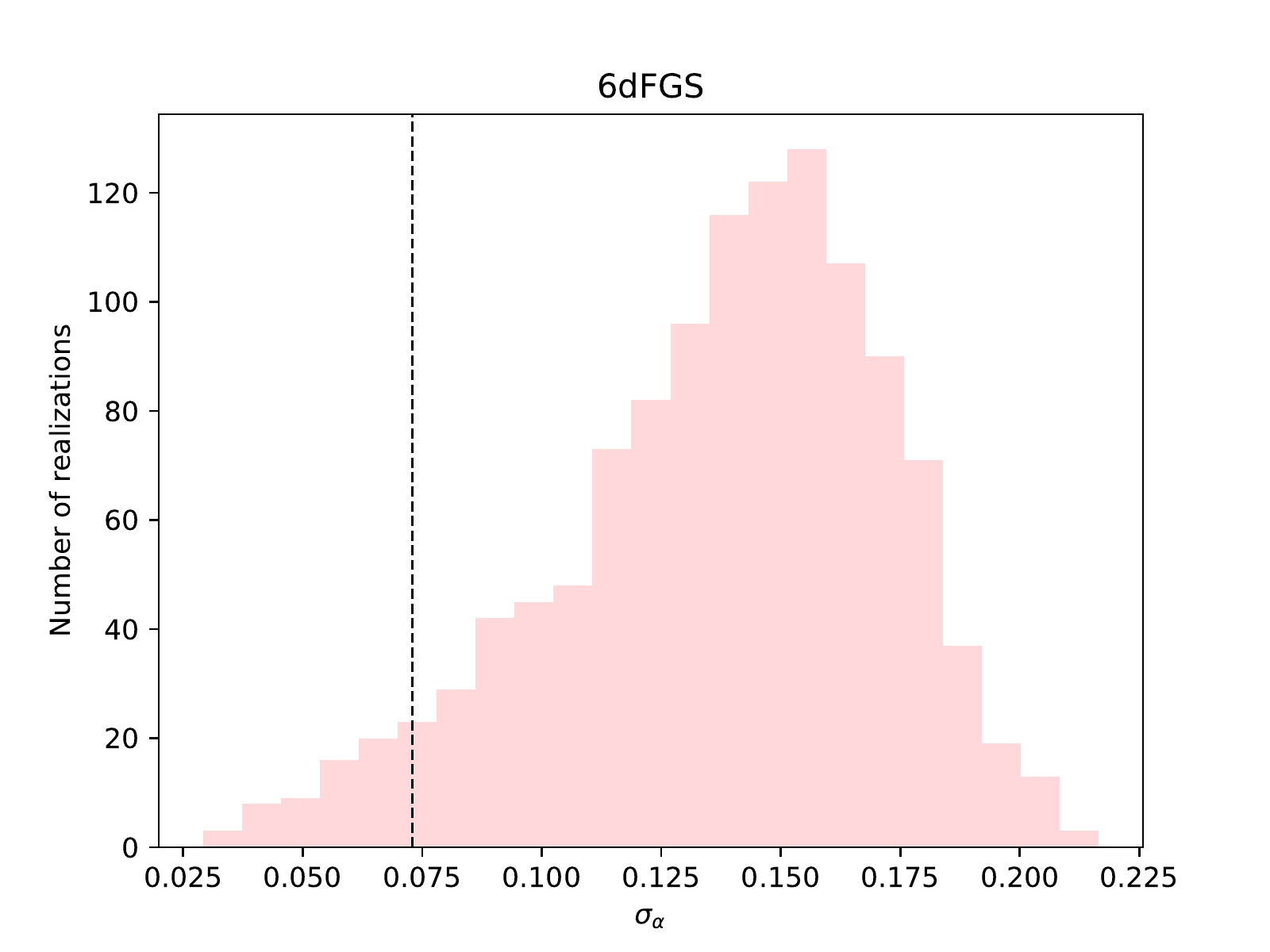}
    \caption{Width of the posterior probability distributions for $\alpha$ for the 6dFGS COLA mocks.  $\sigma_\alpha$ is computed by determining the width of the $68\%$ confidence regions of the posterior of each mock.  The dashed black line displays the variance as inferred from the posterior of the 6dFGS data.}
    \label{fig:6dfsigalpha}
\end{figure}

Figure \ref{fig:6dfsigalpha} displays the distribution of $\sigma_\alpha$, the width of the $68\%$ confidence region of the posterior of each mock.  The dashed black line corresponds to the value of $\sigma_\alpha$ obtained from the posterior of the data.  In contrast to what we found for the WiggleZ Survey, there is a significant difference ($\sim 56\%$) between the value of  $\sigma_\alpha$ estimated from the mean of the distribution and from the posterior of the data.  This result can be understood by inspecting the correlation function of the data, which shows a enhanced peak around $110 \, h^{-1}$ Mpc with respect to the mock mean, as displayed in Figure \ref{fig:6dfcf}.  Due to this noise fluctuation at the BAO scale, we obtain a narrower data posterior and thus a lower value of $\sigma_\alpha$.

\subsection{Baryon Oscillation Spectroscopic Survey (BOSS)}

\subsubsection{Data and mocks}
\label{sec:mock}

The Baryon Oscillation Spectroscopic Survey (BOSS) \citep{2017MNRAS.464.1168R} used the 2.5 metre telescope at Apache Point Observatory in New Mexico to create the largest current galaxy redshift survey, as part of the Sloan Digital Sky Survey (SDSS) collaboration.  We analysed the final data release of BOSS DR12 \citep{2015ApJS..219...12A}, focussing on the CMASS sample of $618{,}806$ Luminous Red Galaxies (LRGs) in the Northern Galactic Cap (NGC) sky region, spanning redshifts $0.43<z<0.7$ and area $7{,}606$ deg$^2$.  Galaxies are weighted to correct for systematics due to fiber collisions, redshift failures, stellar contamination and seeing.  FKP weights are applied in order to maximize the clustering signal with respect to the noise. 

We also analysed 1000 Quick Particle Mesh (QPM) galaxy catalogues corresponding to the BOSS CMASS sample in the NGC region.  Briefly, the mocks are generated by a particle-mesh gravity solver describing the non-linear evolution of the density field, but without sufficient resolution to define dark matter halos.  Particles are selected from the evolved density field based on the local density, approximating the small-scale distribution of halos found in high-resolution simulations.  Galaxies are then assigned to the particles according to a halo occupation distribution (HOD) \citep{2012ApJ...745...16T} which matches the clustering of the BOSS sample as a function of redshift and including observational effects such as fibre collisions.  The fiducial cosmology of the mocks is a flat $\Lambda$CDM Universe with $\Omega_m = 0.29$, $\Omega_b h^2 = 0.0225 $, $\Omega_\Lambda = 0.71$, $h = 0.7$, $n = 0.97$ and $\sigma_8 = 0.8$.  Full details on how the mocks are generated  can be found in \cite{2014MNRAS.437.2594W}.

We considered two different analyses: the first uses six equal-volume subsets of the CMASS NGC sample, where each sub-sample contains $\sim 100{,}000$ galaxies.  The second analysis considers instead the full CMASS NGC sample.  These analyses of the sub-samples and full sample of the same survey allow us to study how the noise component in the data posterior depends specifically on the survey volume.

\subsubsection{Analysis of sub-samples}
\label{subsec:bosssub}

We first present the results derived from the 6 sub-volumes defined as described in Table \ref{tab:sub}.  Owing to the high number density of the CMASS catalogue, the BAO signal is typically detectable in each individual sub-volume, and every sub-sample can be interpreted as an approximately independent realization of the Universe.

% \begin{table}
\begin{table*}
%\begin{minipage}[t]{0.48\linewidth}  
\begin{tabular}{ccccccc}
      &1&2&3&4&5&6\\
        RA [deg] & 108 ; 160 & 160 ; 200 & 200 ; 265  & 108 ; 160   & 160; 205 & 205 ; 265\\
        Dec [deg] & -4 ; 27 & -4 ; 27 &  -4 ; 27 & 27; 70 & 27; 70 & 27;70 \\
        $\mathrm N^\circ $ & 95730 &  97603 & 117018 &  95638 & 104470 & 108288
    \end{tabular}
  \caption{Subsamples obtained by dividing the area of the BOSS DR12 CMASS NGC catalogue into 6 RA and Dec intervals, in the redshift range $0.43 - 0.7$.  The selected subsamples contain a comparable number of objects.}
    \label{tab:sub}
%\end{minipage}
%\end{table}
\end{table*}
 
The following analysis is repeated for each of the $6000$ sub-mocks and the 6 corresponding subsets of the data catalogue.  We compute the monopole correlation function $\xi^\mathrm{sub}$ with  ($sub=1\dots 6$) as described in Sec. \ref{sec:corrfuncmeas}, in the separation range $50 < s < 180 \, h^{-1}$ Mpc using $5 \, h^{-1}$ Mpc bins.  FKP weights with $P_0 = 10^4 \, h^{-3}$ Mpc$^3$ are applied to individual galaxies together with systematic weights described in \citet{2015ApJS..219...12A}.  The covariance matrix of the correlation function in each sub-volume is a $26 \times 26$ matrix,
\begin{equation}\begin{split}
C^{\mathrm{sub}}_{(i,j)} = \frac{1}{N - 1 } \sum_{n=1}^{N} \big[ \xi^{\mathrm{sub}}_n(s_i) - \hat{\xi}^\mathrm{sub}(s_i) \big]  \times \big[  \xi^{\mathrm{sub}}_n(s_j) - \hat{\xi}^\mathrm{sub}(s_j) \big],
\label{covmatsub}
\end{split}\end{equation}
where $\hat{\xi}$ denotes the mean correlation function of $N = 1000$ sub-mock realizations.

We fit for the BAO peak position in the correlation functions in the same manner as described in Sec. \ref{subsec:wigglefit} and \ref{subsec:6dfit}.  We select the same fiducial cosmology as for the QPM mocks and run the MCMC pipeline with a prior $0.8 < \alpha < 1.2 $ and flat, wide priors  on the nuisance parameters $a_1$, $a_2$, $a_3$, $B$.

\begin{figure*}\label{subsam1}
\begin{minipage}{19cm}
\centering
 % \begin{multicols}[3]
    \includegraphics[scale=0.33]{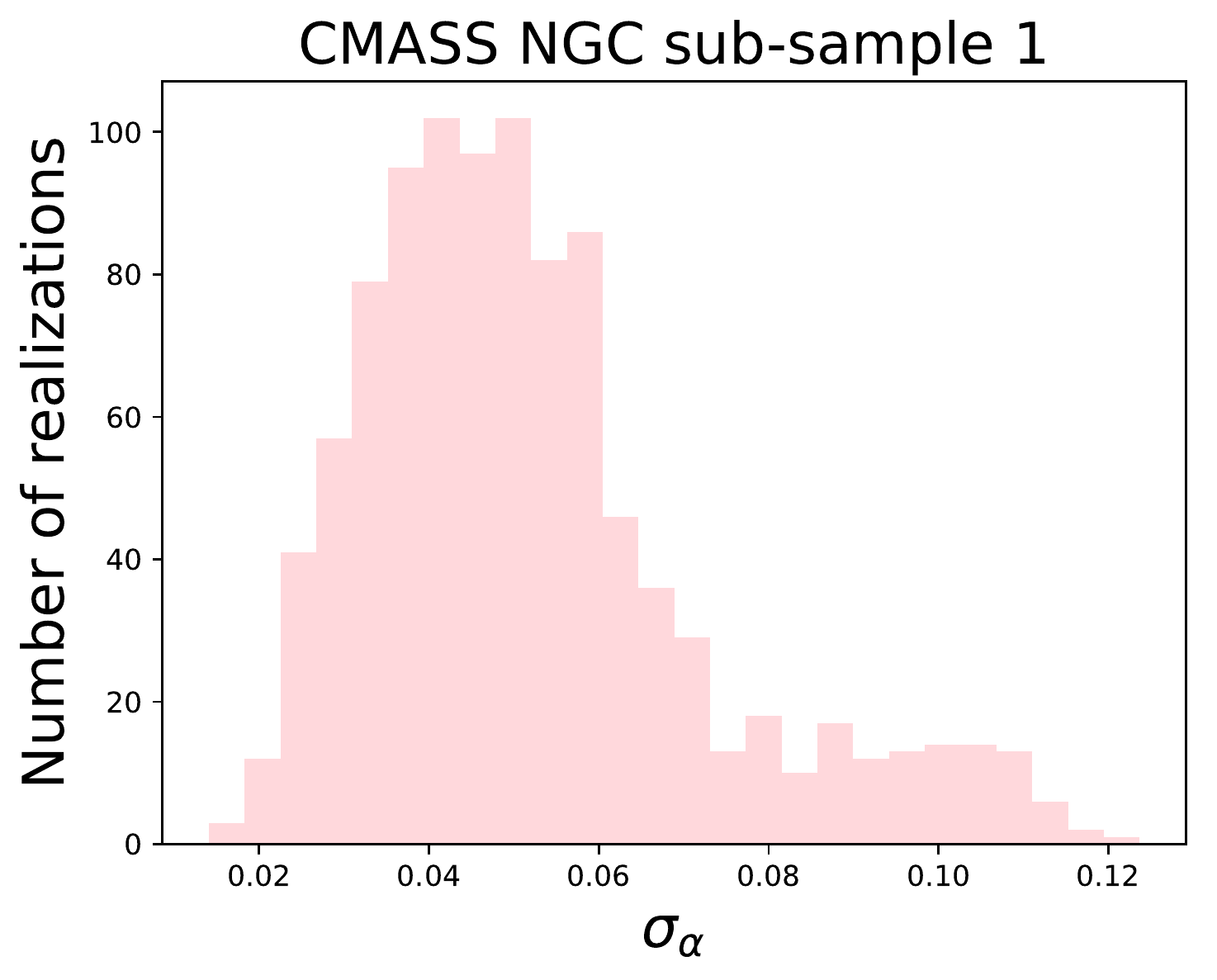}
    \includegraphics[scale=0.33]{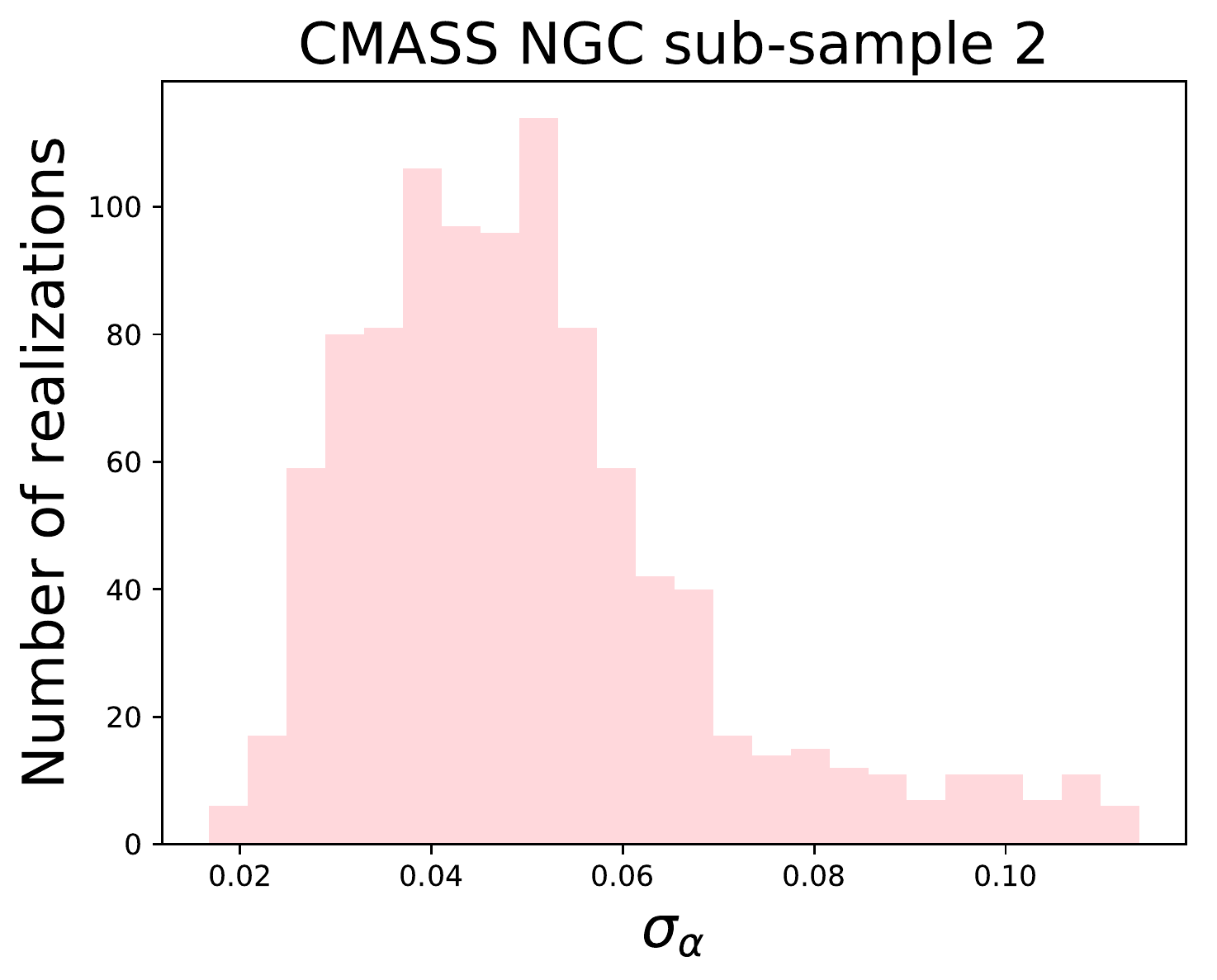}
    \includegraphics[scale=0.33]{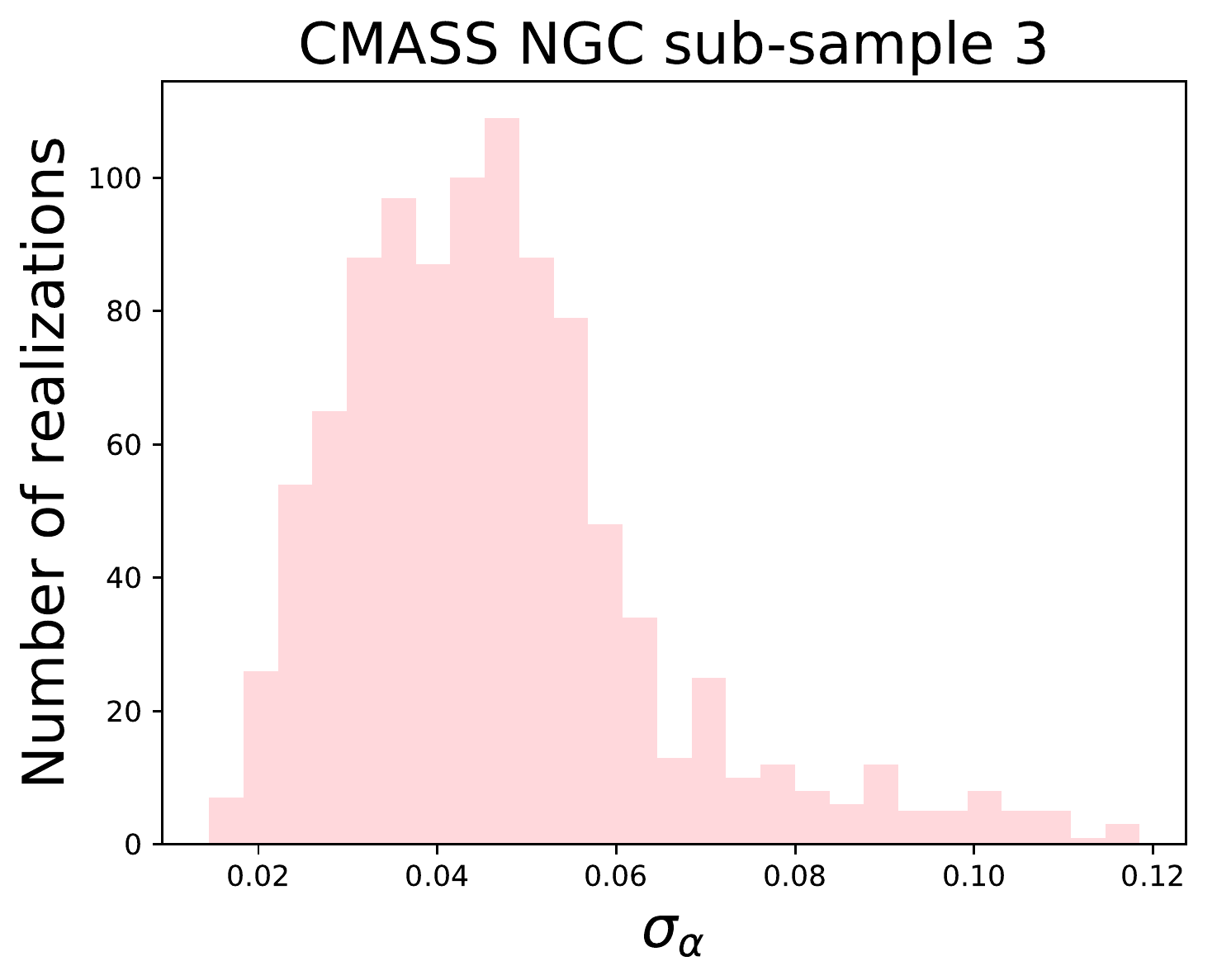}
    \includegraphics[scale=0.33]{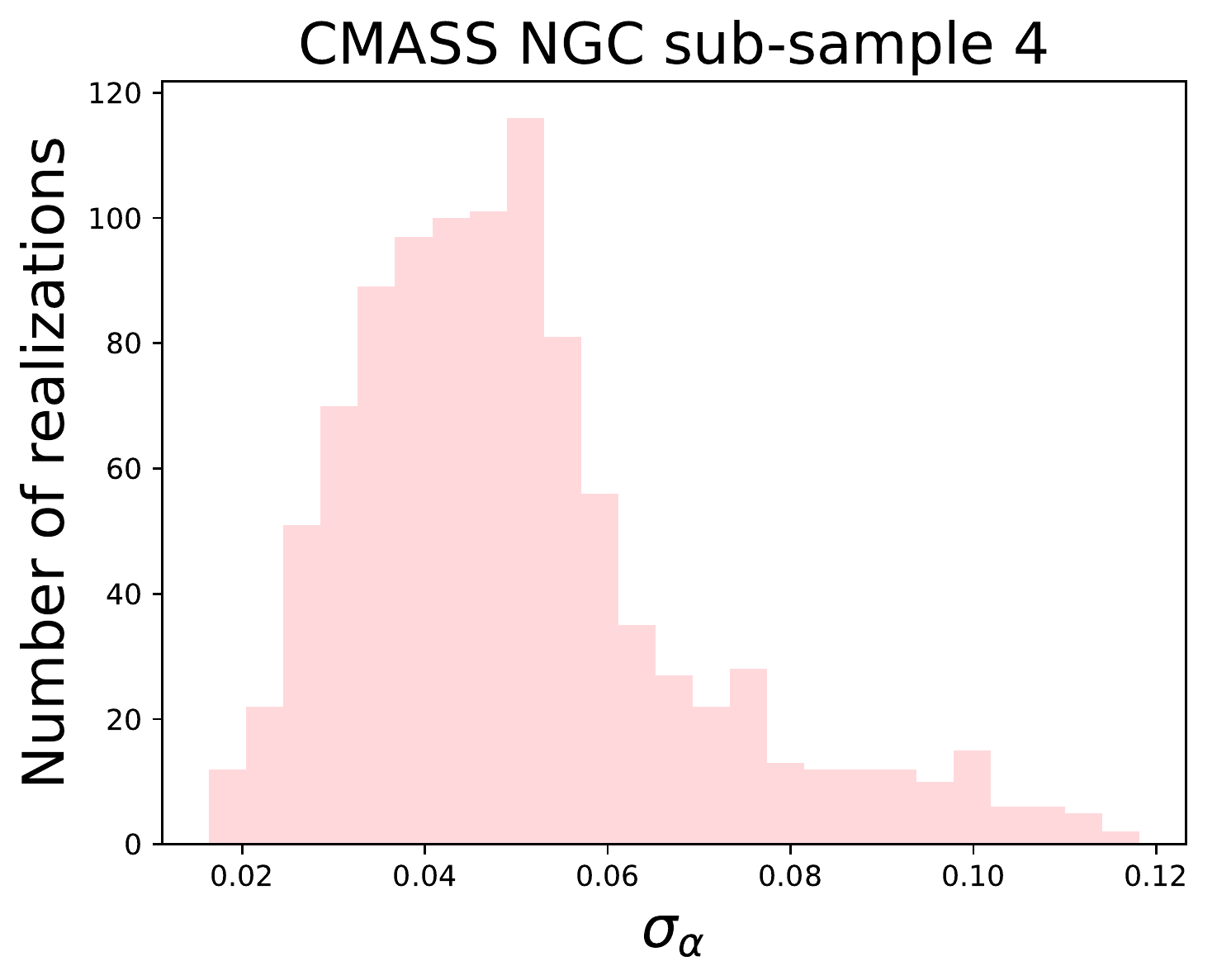}
    \includegraphics[scale=0.33]{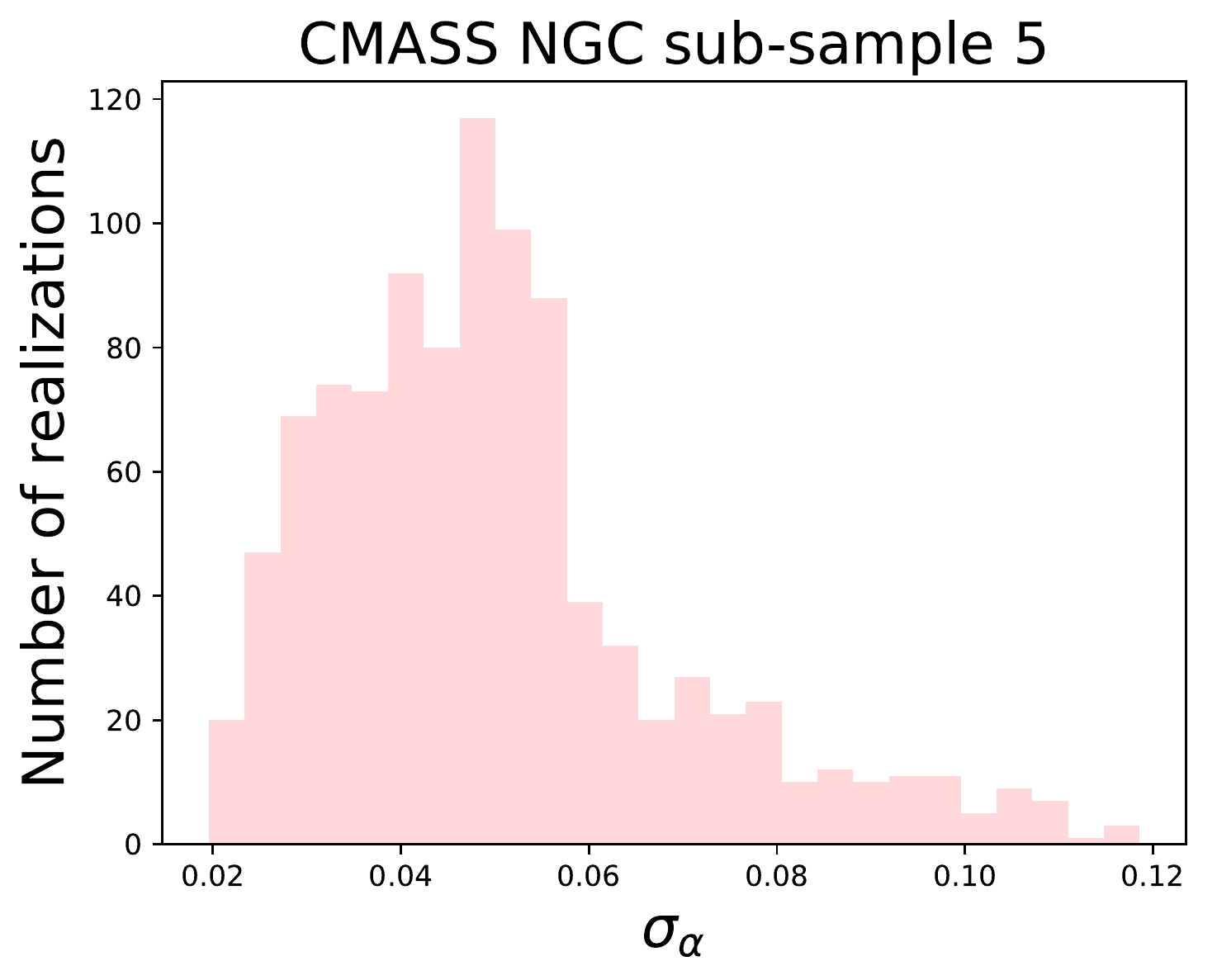}
    \includegraphics[scale=0.33]{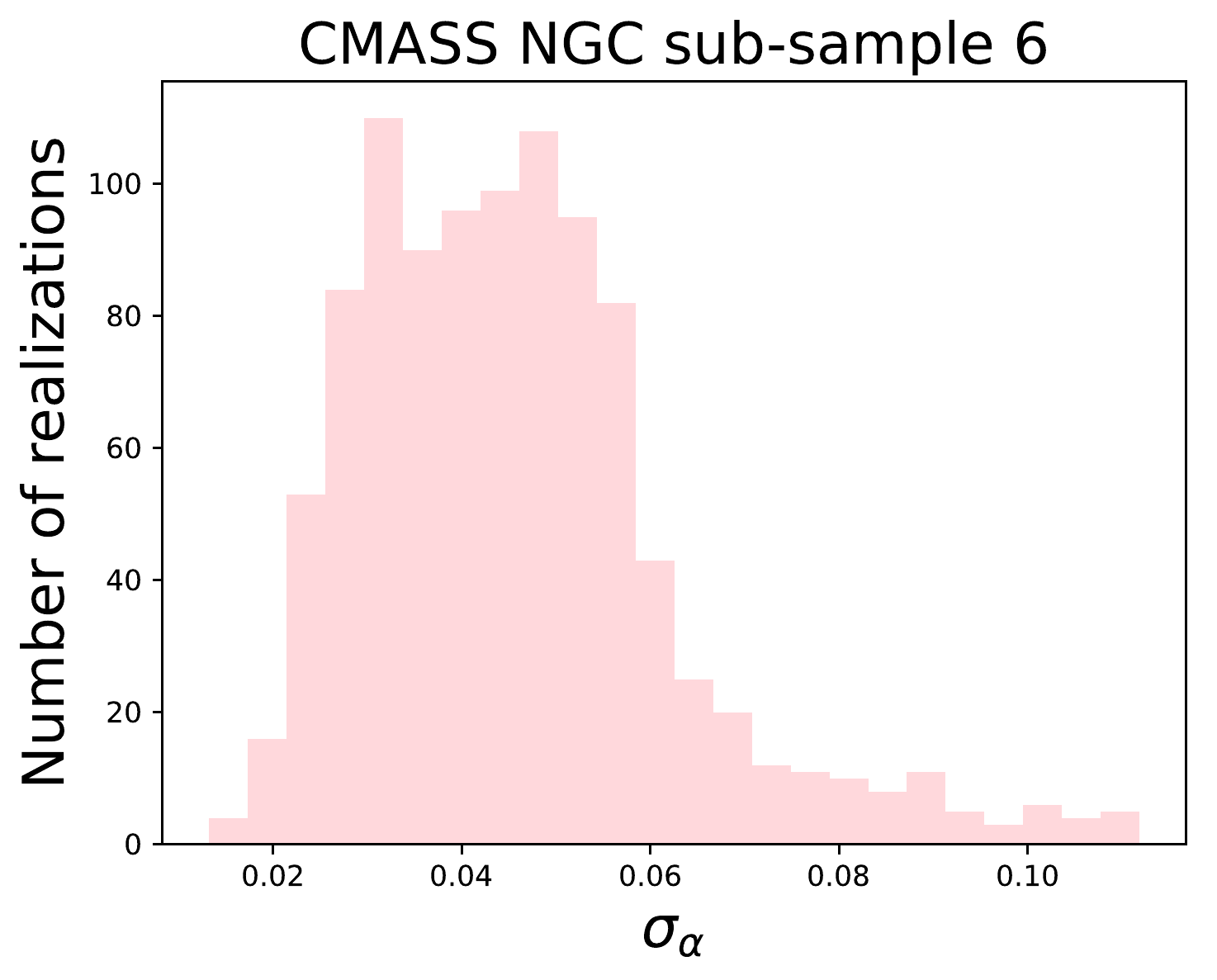}
 %\end{multicols}
    \end{minipage}
        \caption{Width of the posterior probability distributions for $\alpha$ for 6 equal sub-samples of mocks corresponding to the BOSS CMASS NGC sample.  Each value of   $\sigma_\alpha$ is given by the width of the $68\%$ confidence region of the posterior distribution measured for each realization. 
        %\cab{[Suggest re-formatting figure using larger font size if possible, currently difficult to read labels]}
        }
    \label{fig:samplesvar}
\end{figure*}

\begin{figure}
%\begin{minipage}{19cm}
\centering
    \includegraphics[width =0.92\columnwidth]{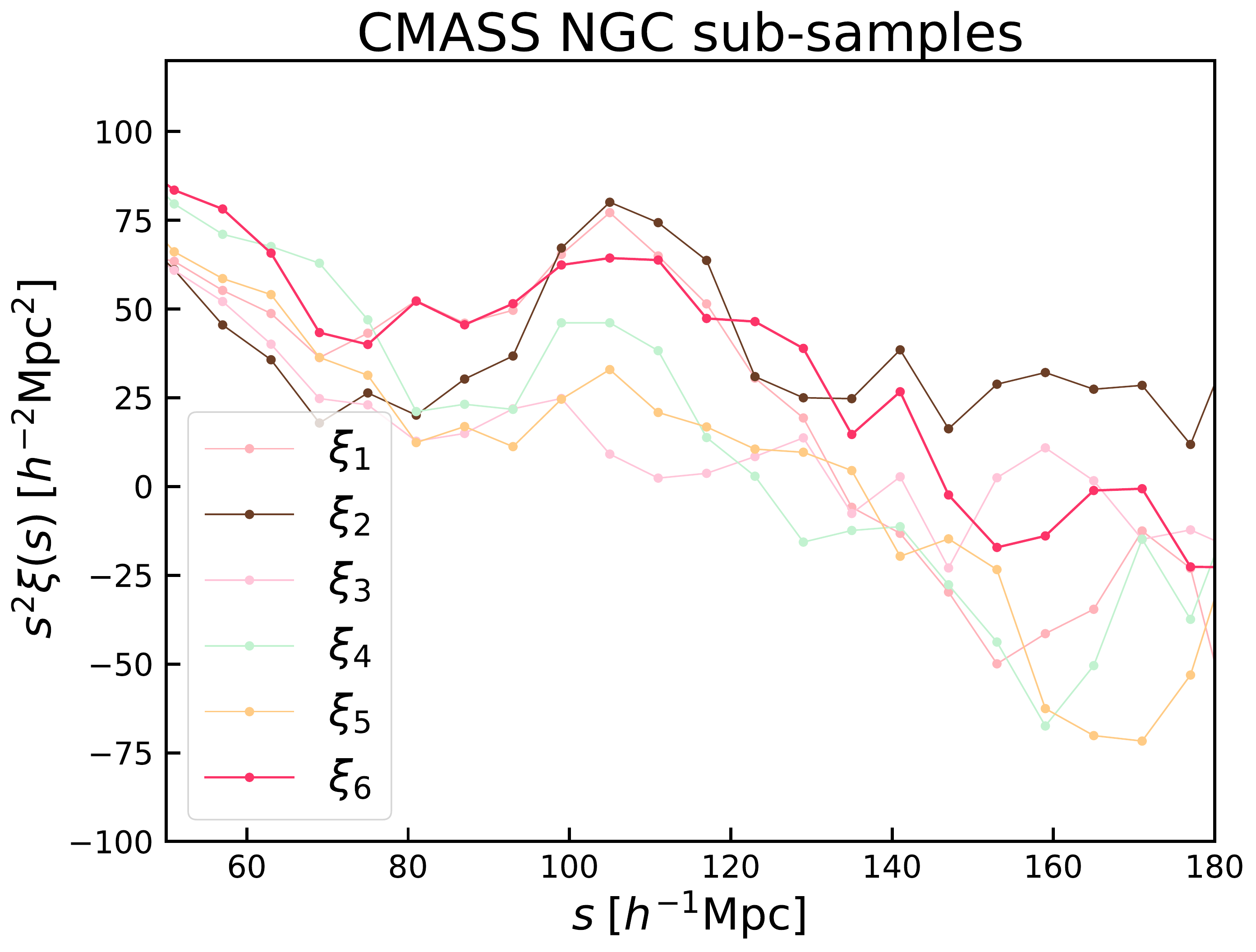}
    \includegraphics[width =0.93\columnwidth]{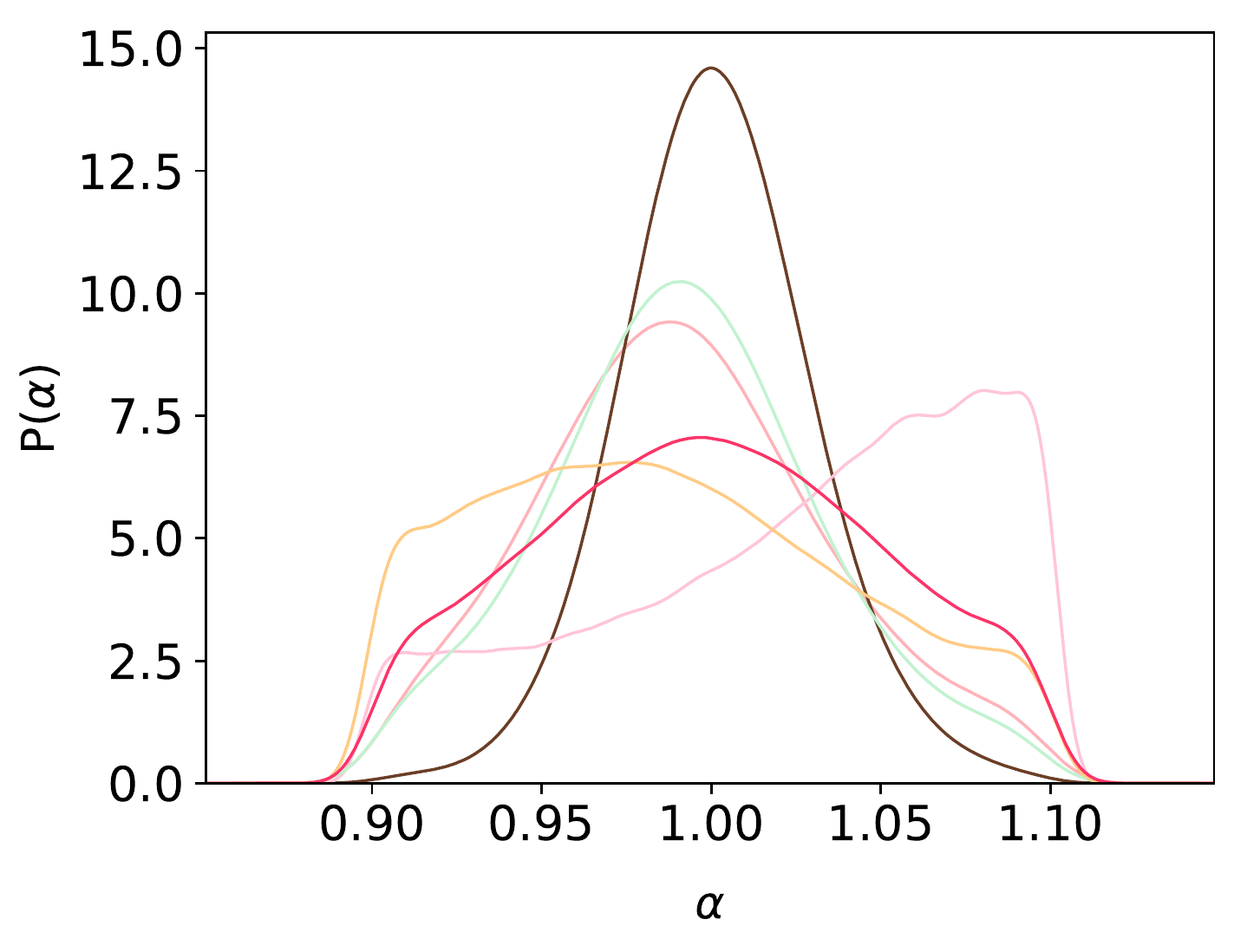}  %  
    \caption{Top panel: the correlation functions of  the 6 sub-samples for one QPM mock realization of the  CMASS sample. Bottom panel: the normalized posterior probability distributions of $\alpha$ obtained by fitting the BAO peak in each sub-sample. (Colours and labels correspond to the same sub-sample between the two panels).
    %\cab{[Suggest adding title to upper panel ``CMASS NGC sub-samples'' and adding $y$-axis scale to lower panel]}
    }
    \label{fig:samplesvar2}
\end{figure}

Figure \ref{fig:samplesvar} displays the distribution of the $68\%$ confidence ranges in the posterior probability distribution, $\sigma_\alpha$, obtained for the 1000 mocks in the 6 sub-samples considered.  It is clear from the panels that the distribution of $\sigma_\alpha$ is very similar in each sub-sample, given the homogeneity of the BOSS sample.  We note that in each sub-sample the distribution of  $\sigma_\alpha$ is not symmetric, and varies over a wide range $0.02 - 0.12$.  As we discussed in Sec. \ref{sec:error}, this is due to the noise in the posterior probability distribution related to sample variance fluctuations.  As a further illustration, in Fig. \ref{fig:samplesvar2} we select one mock configuration and plot the 6 subset correlation functions (top panel) and the relative posteriors of $\alpha$ obtained from the fit to each subset (bottom panel).  We note that narrower posteriors correspond to narrower/enhanced peaks in the galaxy correlation function, as illustrated by sub-sample \#2 (the brown line in the top and bottom panels). 

\subsubsection{Analysis of full sample}\label{subsec:bossfull}

\begin{figure}
      \centering
     \includegraphics[width= \columnwidth]{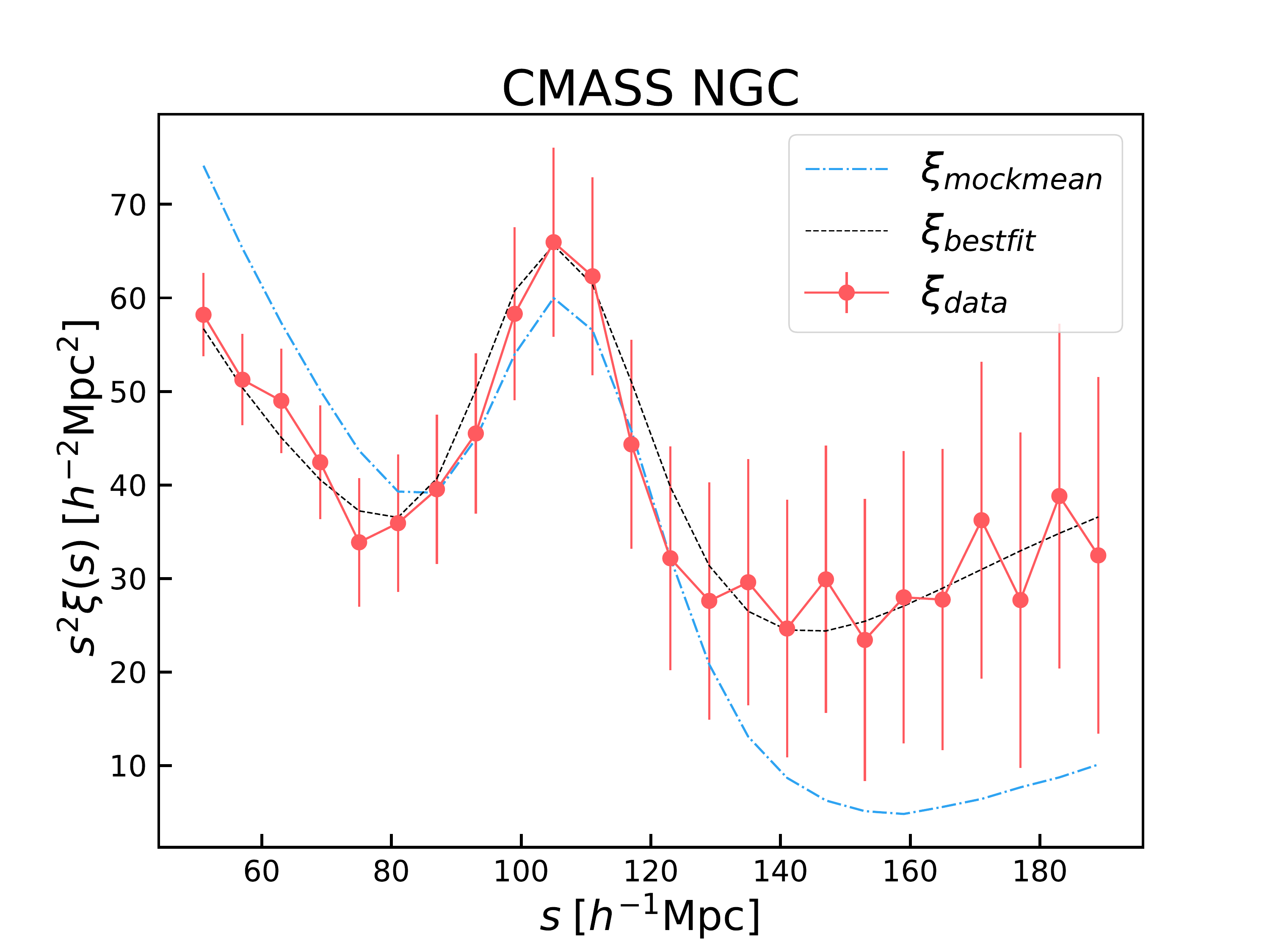}
\caption{The correlation function of the BOSS CMASS NGP galaxy sample.  Red solid line: the correlation function of the data for scales between $50-180 \, h^{-1}$ Mpc in 26 bins. The blue dotted line corresponds to the mean of $1000$ QPM mock correlation functions. The black dashed line is the best-fitting model for $\xi$.}
  \label{fig:bosscf}
 \end{figure}

\begin{figure}
    \centering
        \includegraphics[width= \columnwidth]{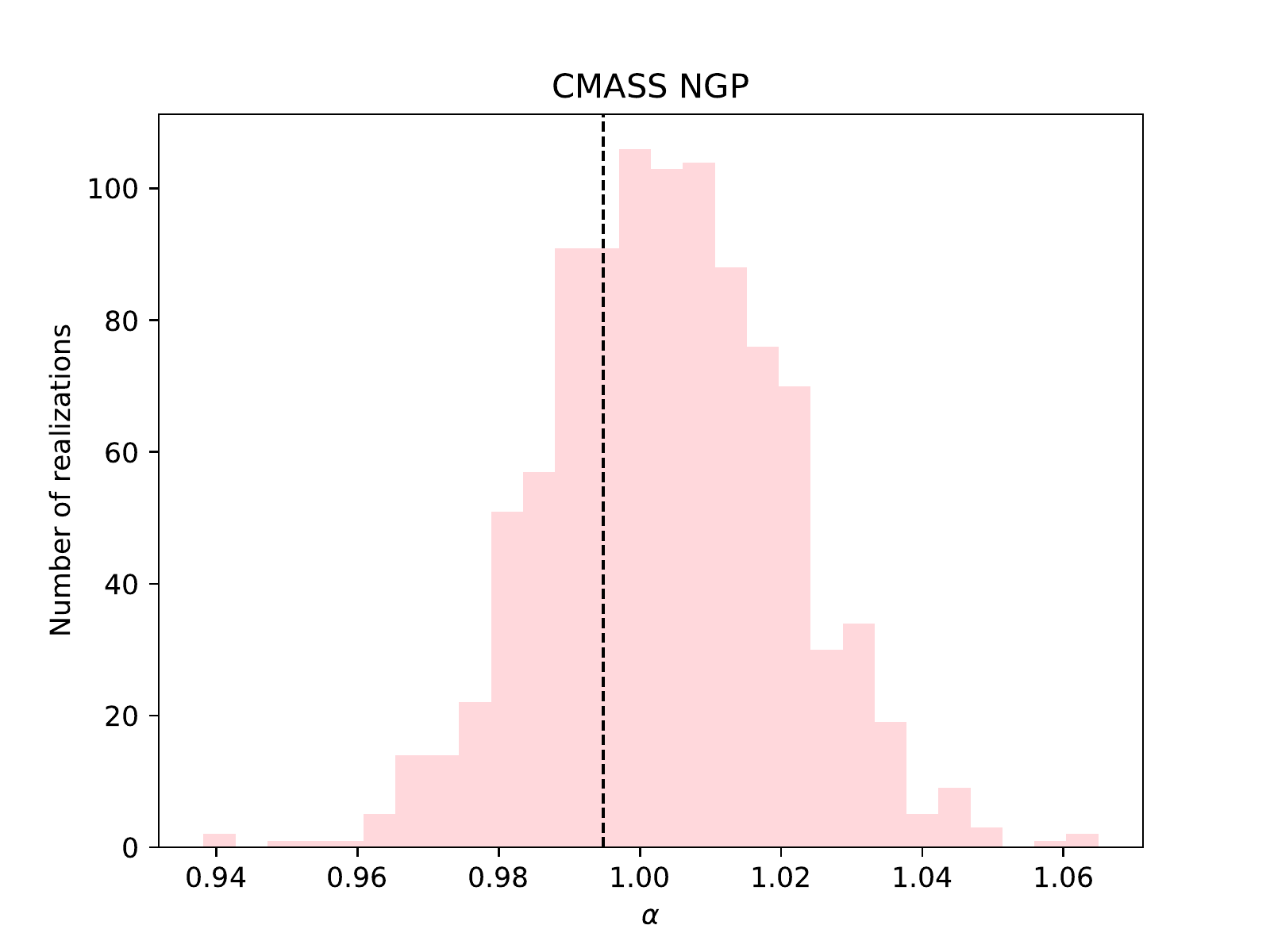}
    \caption{Best-fitting BAO distortion parameters $\alpha$ for the CMASS NGP QPM mocks. Each $\alpha$ is computed from  the mean of the posterior distribution obtained for each mock.  The dashed black line displays the mean value inferred from the data posterior.}
    \label{fig:bossalpha}
\end{figure}

\begin{figure}
    \centering
        \includegraphics[width= \columnwidth]{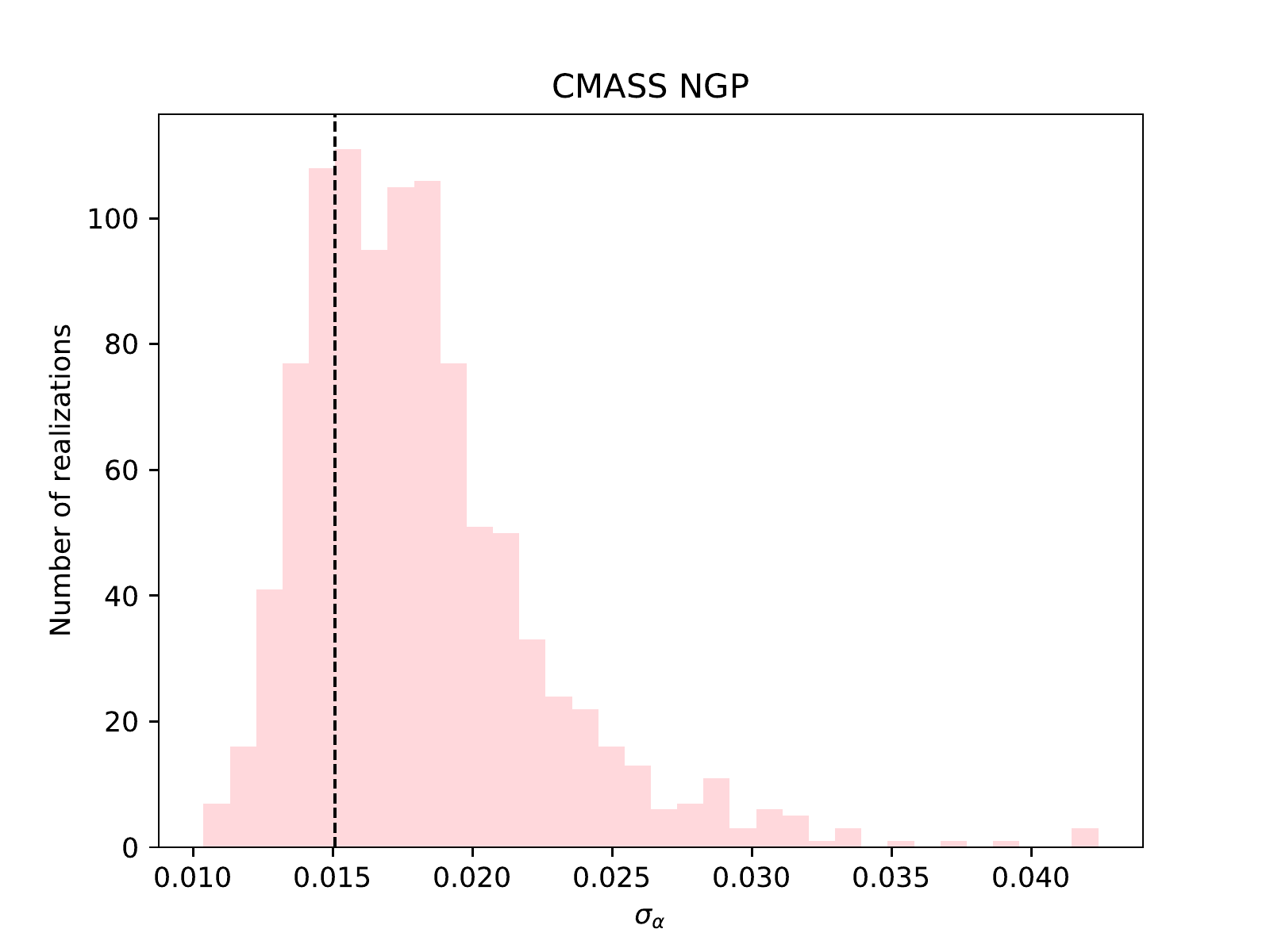}
 \caption{Width of the posterior probability distribution for $\alpha$ for the 1000 CMASS NGP QPM mocks.  Each $\sigma_\alpha$ corresponds to the width of the $68\%$ confidence regions of the posterior of each mock.  The dashed black line displays the variance inferred from the data posterior.}
    \label{fig:bosssigmaalpha}
\end{figure}

We repeated the analysis described in Sec. \ref{subsec:bosssub} for the full CMASS sample in the NGC region.  Figure \ref{fig:bosscf} displays the correlation function measured from the CMASS NGC catalogue between 50 and 180 $h^{-1}$ Mpc, compared with the mean of the 1000 QPM mocks (blue dot-dashed line) and the best-fitting model (black dashed line).  Similarly to the case of 6dFGS, the correlation function measured from the data features an enhanced peak compared to the average of the mocks.  Figures \ref{fig:bossalpha} and \ref{fig:bosssigmaalpha} show the distribution across 1000 mocks of $\alpha$ and the error $\sigma_\alpha$ inferred from the individual posterior of each mock.  In Table \ref{tab:resultALL} we list the results obtained fitting the  CMASS sample:  $\alpha$ and $\sigma_\alpha$ are the mean and width of the posterior distribution of the data. The reduced chi-squared statistic is obtained using $D.O.F. = (24-5)$.

%BOSS FULL DR12


\subsection{Extended Baryon Oscillation Spectroscopic Survey (eBOSS)}
\subsubsection{Data and mocks}

The 6-year extended Baryon Oscillation Spectroscopy Survey (eBOSS) is a component of SDSS-IV \citep{2016AJ....151...44D} which started in July 2014 and is mapping the largest volume to date of any cosmological redshift survey. The observations target multiple tracers and will include more than $250{,}000$ luminous red galaxies (LRGs) and $195{,}000$ emission line galaxies (ELGs) at effective redshifts $z = 0.72$ and $0.87$, and over $500{,}000$ quasars in the redshift range $0.8 < z < 2.2$.  For this analysis we make use of the latest available eBOSS catalogue, the QSO DR14 sample.  We use $97{,}456$ quasars from the eBOSS DR14 NGC catalogue selected in the redshift range $0.8 < z < 2.2$.  We also make use of 400 corresponding QPM mock catalogues, produced as described in Sec. \ref{sec:mock}, where galaxies are  assigned according to an HOD \citep{2012ApJ...745...16T}.  These mocks are based on 100 independent boxes with side  $ 5120 \, h^{-1} \mathrm{Mpc}$, re-mapped to match the observed data catalogue using the MAKE-SURVEY code \citep{2010ApJS..190..311C}.  The fiducial cosmology of the mocks is a flat $\Lambda$CDM model with $\Omega_m = 0.31$, $\Omega_b h^2 = 0.022$, $h = 0.676$ and $m_\nu = 0.06$ eV.

\subsubsection{Analysis}

\begin{figure}
      \centering
     \includegraphics[width= \columnwidth]{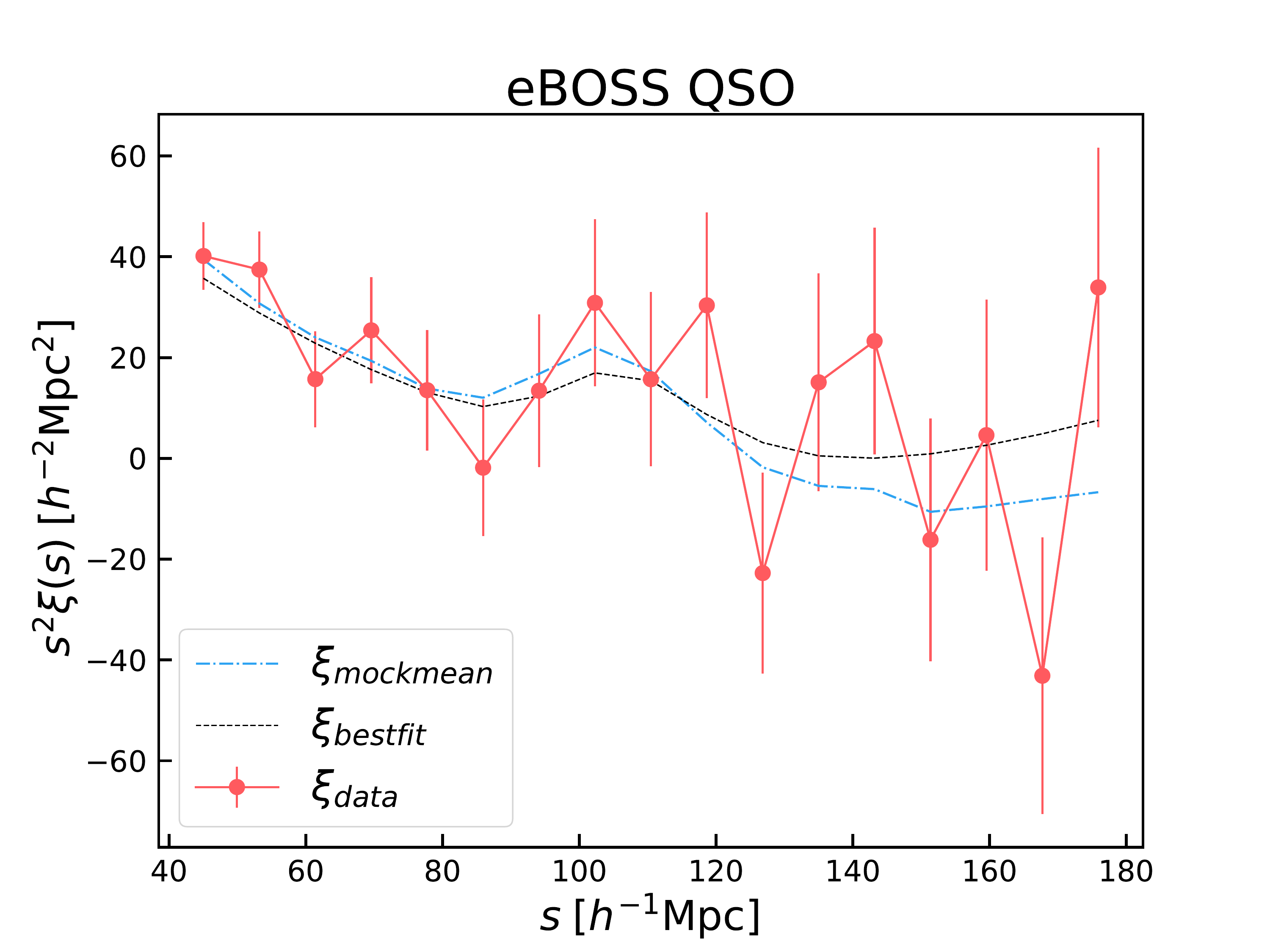}
\caption{The correlation function of the eBOSS QSO sample.  Red solid line: the correlation function of the QSO data at scales between $50-180 \, h^{-1}$ Mpc in 26 bins.  The blue dotted line corresponds to the mean of $400$ QPM mock correlation functions.  The black dashed line is the best-fitting model for $\xi$.}
  \label{fig:ebosscf}
 \end{figure}

We measured the correlation functions of the data and mock samples with the estimator described in Sec. \ref{sec:corrfuncmeas}.  Galaxy weights are applied, including FKP weights with $P_0  = 6 \times 10^3 \, h^{-3} \mathrm{Mpc}^3$ and completeness weights.  Full details about the weights applied to the data to correct for systematics are described in \citet{2018MNRAS.473.4773A}.  Figure \ref{fig:ebosscf} displays the correlation function measured from the eBOSS NGC catalogue between 50 and 180 $h^{-1}$ Mpc  with $5 \, h^{-1}$ Mpc bins (red solid line), compared with the mean of the 400 QPM mocks (blue dot-dashed line) and the best-fit model obtained fitting the data (black dashed line). 
 
We performed an isotropic BAO fit to each of the 400 QPM mock realizations and the NGC data catalogue using the templates described in Sec. \ref{sec:isotropicfit}.  We used the MCMC pipeline introduced in Sec. \ref{subsec:wigglefit}, using a prior $0.8 < \alpha < 1.2$ and flat, wide priors on $a_1$, $a_2$, $a_3$, $B$.  $\Sigma_{nl}$ is fixed to its best-fit value for the mock mean, $5.2 \, h^{-1}$ Mpc.  Table \ref{tab:resultALL} displays the results obtained fitting the eBOSS QSO data, with $\alpha$ and $\sigma_\alpha$ estimated as the mean and width of the data posterior distribution. The $D.O.F.$ in the reduced $\chi^2$ is given by $22$ bins - $5$ free parameters.

\begin{figure}
    \centering
        \includegraphics[width= \columnwidth]{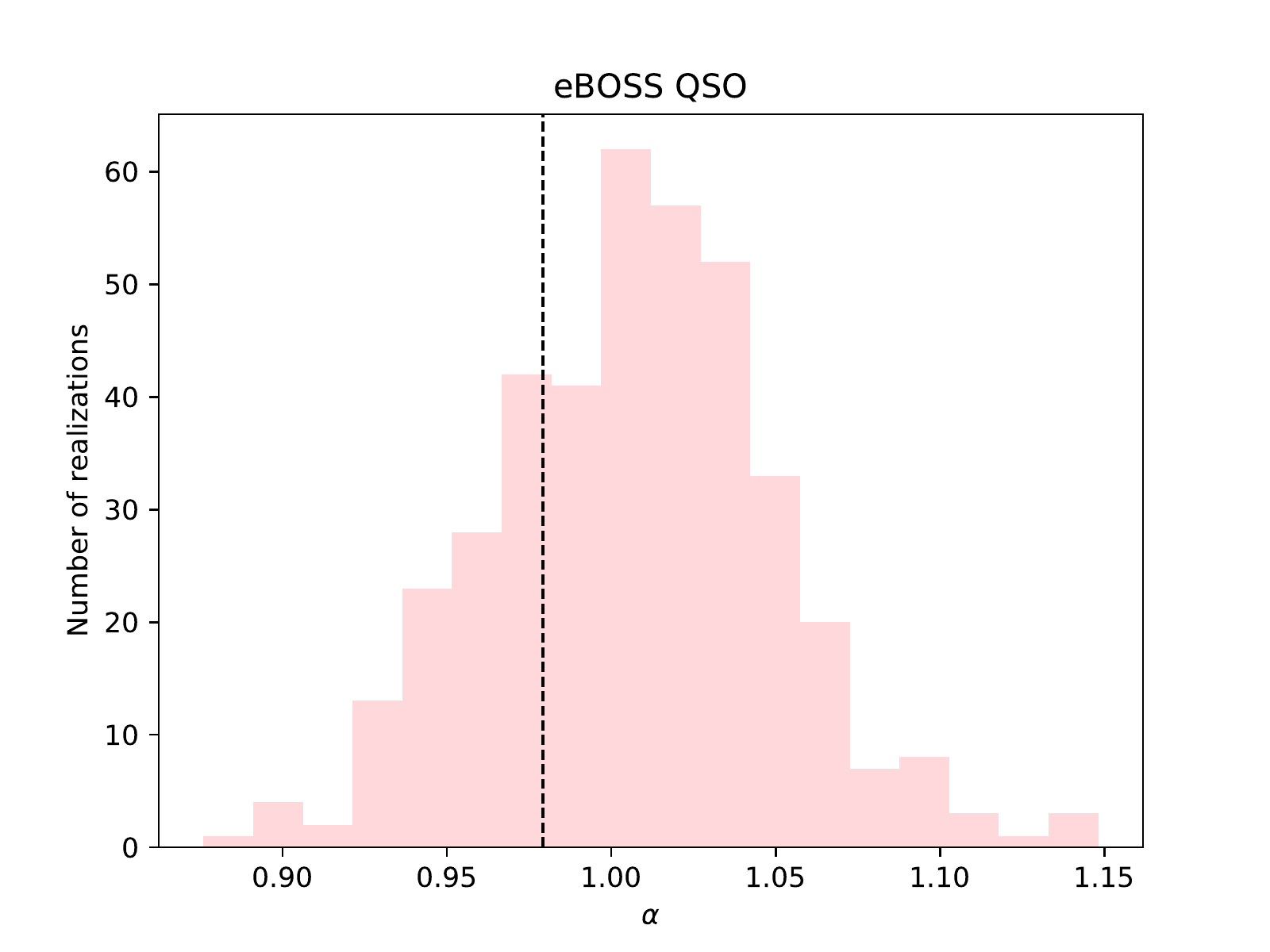}
    \caption{Best-fitting BAO distortion parameter $\alpha$ for the 400 eBOSS QSO QPM mocks; each $\alpha$ is computed from  the mean of the posterior distribution obtained for each mock. The dashed black line displays the mean value inferred from the data posterior.}    \label{fig:eboss1}
\end{figure}

\begin{figure}
    \centering
        \includegraphics[width= \columnwidth]{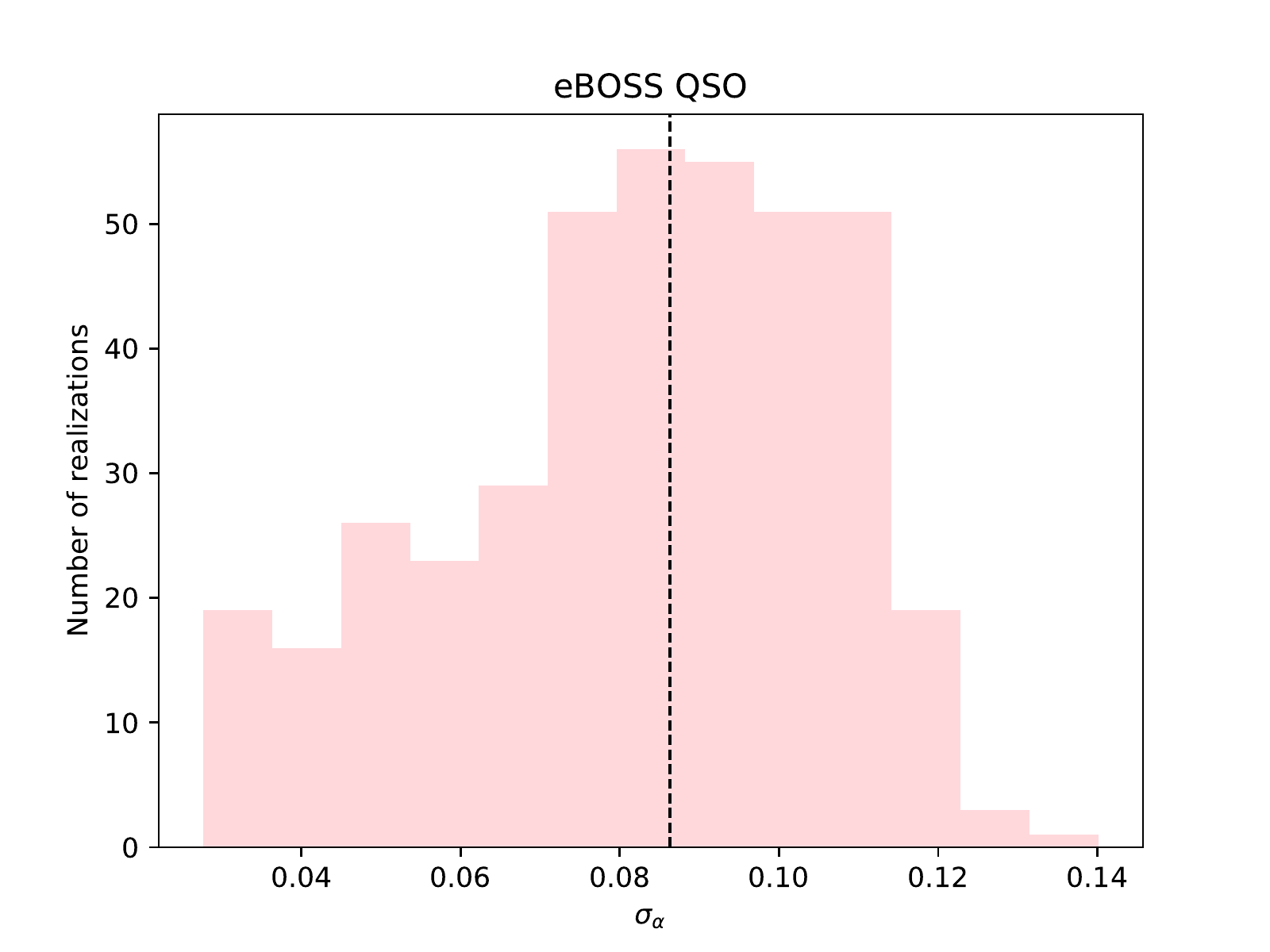}
    \caption{Width of the posterior probability of $\alpha$ for the 400 eBOSS QSO QPM mocks; each $\sigma_\alpha$ corresponds to the width of the $68\%$ confidence regions of the posterior of  each mock.  The dashed black line displays the variance inferred from the data posterior.}
    \label{fig:eboss2}
\end{figure}

In Figures \ref{fig:eboss1} and \ref{fig:eboss2}, as for the other surveys, we display the distribution of $\alpha$ and $\sigma_\alpha$ measured from the individual posterior distributions.  The corresponding data measurements are indicated with dashed black lines.  We note that the distribution of the $\sigma_\alpha$ is not symmetric, similarly to Fig. \ref{fig:samplesvar}.

\section{Fisher matrix forecast}\label{sec:fish2}

As introduced in Sec.\ \ref{sec:fisher}, a Fisher matrix forecast provides a method to propagate the measurement errors of an observed statistic to the minimum possible determination of the error in a fitted parameter, assuming Gaussian statistics and optimal weighting.  This approach yields a useful comparison with the BAO scale errors inferred from the data posterior (which is affected by the noise in a single realization) and the statistical ensemble of mocks (which incorporates the non-Gaussian likelihood surfaces).

We used the method of \cite{2007ApJ...665...14S} to forecast the errors in the isotropic BAO dilation scale $\alpha$ of the galaxy redshift surveys considered in Sec.\ \ref{sec:data}, arising from the oscillatory component of the galaxy power spectrum (see \cite{2014JCAP...05..023F} for a detailed discussion of BAO Fisher matrix forecasts).  We summed the Fisher matrix in redshift slices, using the observed redshift distribution of each survey to determine the shot noise component of the power spectrum error at each redshift.  The other assumptions of our survey models are listed as follows:
\begin{itemize}
\item {\bf WiggleZ:} redshift ranges $0.2 < z < 0.6$, $0.4 < z < 0.8$ 
and $0.6 < z < 1.0$, bias factor $b = 1.2$ and survey area 816 deg$^2$.
\item {\bf 6dFGS:} redshift range $0 < z < 0.2$, bias factor $b = 1.5$ 
and survey area $15{,}419$ deg$^2$.
\item {\bf BOSS CMASS:} redshift range $0.43 < z < 0.7$, bias factor $b 
= 1.9$ and survey area $7{,}606$ deg$^2$.
\item {\bf eBOSS QSO:} redshift range $0.8 < z < 2.2$, bias factor $b 
=  2.35$ and survey area $1{,}215$ deg$^2$.
\end{itemize}
Following \cite{2007ApJ...665...14S}, we assumed r.m.s. radial displacements across and along the line-of-sight of $\sigma_\perp = 0.758 \, \sigma_0 \, D(z)$ and $\sigma_\parallel = 0.758 \, \sigma_0 \, D(z) \, (1+f)$ where $\sigma_0 = 12.4 \, (\sigma_8/0.9) \, h^{-1}$ Mpc, $D(z)$ is the growth factor normalized to 1 at $z=0$, and $f$ is the growth rate of structure in the redshift slice.  We therefore do not include density-field reconstruction (which would reduce these displacements) in our model.  Our forecasts are compared to the other methods of error determination in the following section. 

\section{Discussion}\label{sec:discussion}

\begin{figure*}
\begin{minipage}{19cm}
\centering
 % \begin{multicols}[3]
     \includegraphics[scale=0.5]{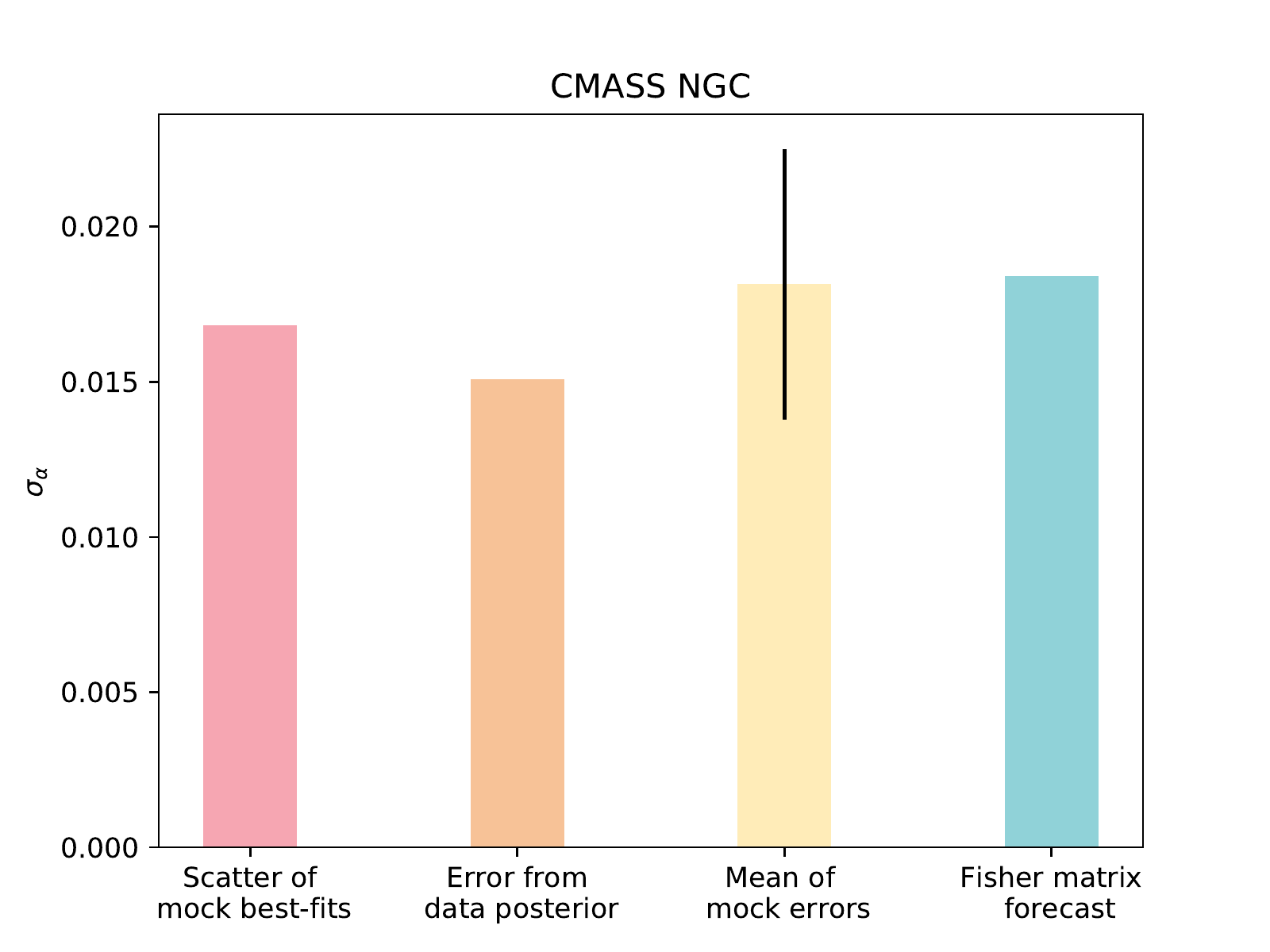}
          \includegraphics[scale =0.5]{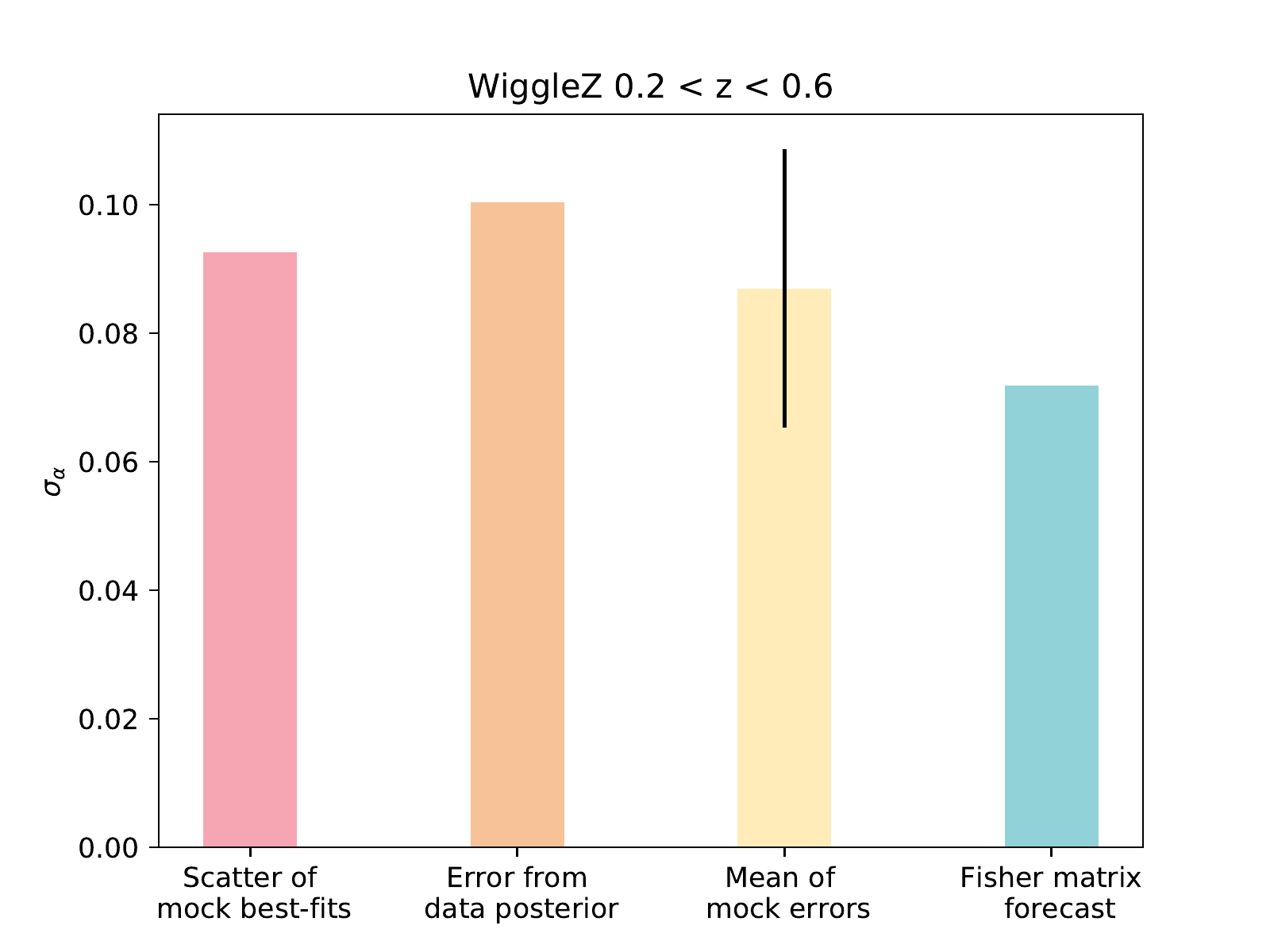}

     \includegraphics[scale =0.5]{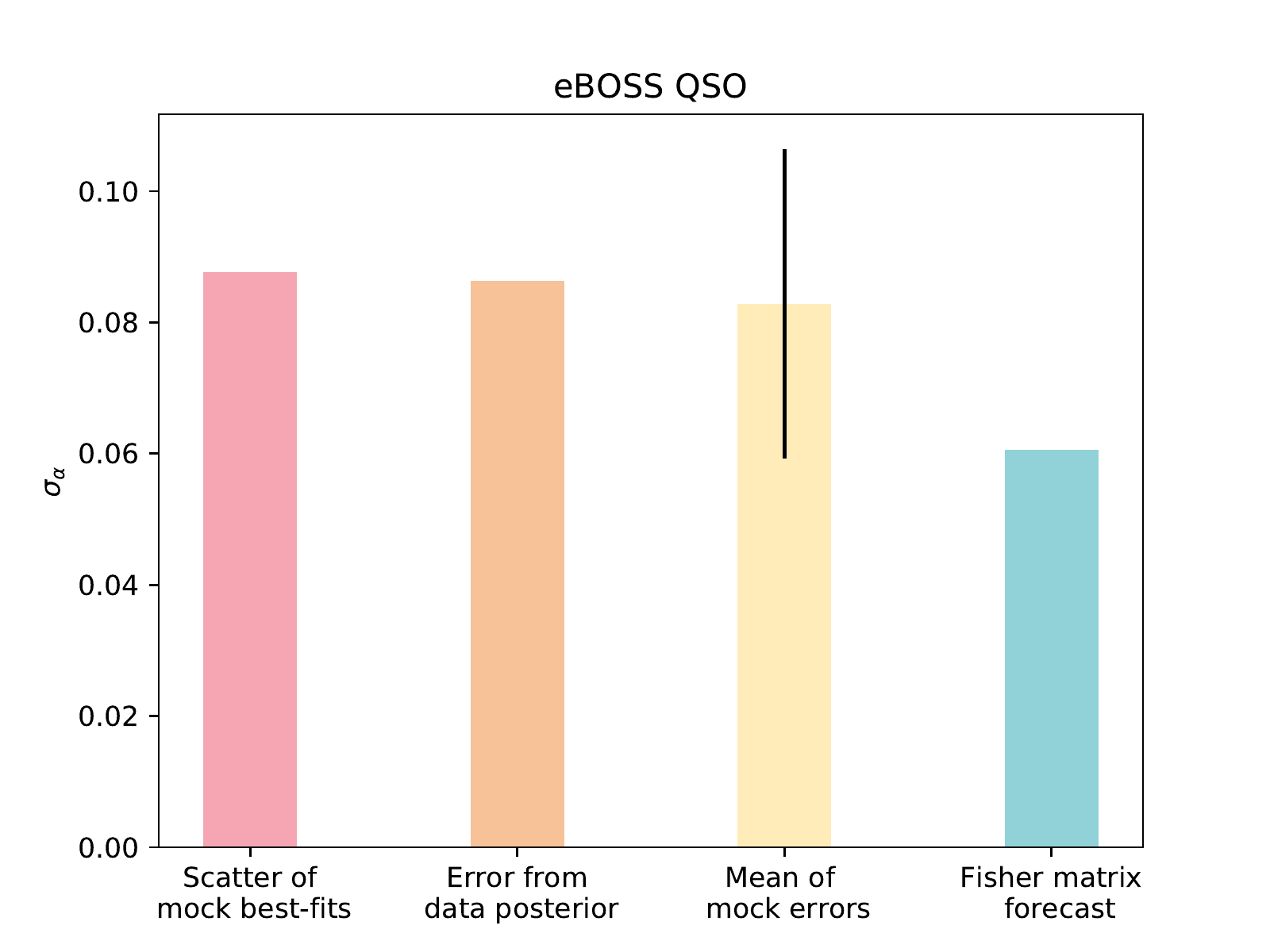}
          \includegraphics[scale =0.5]{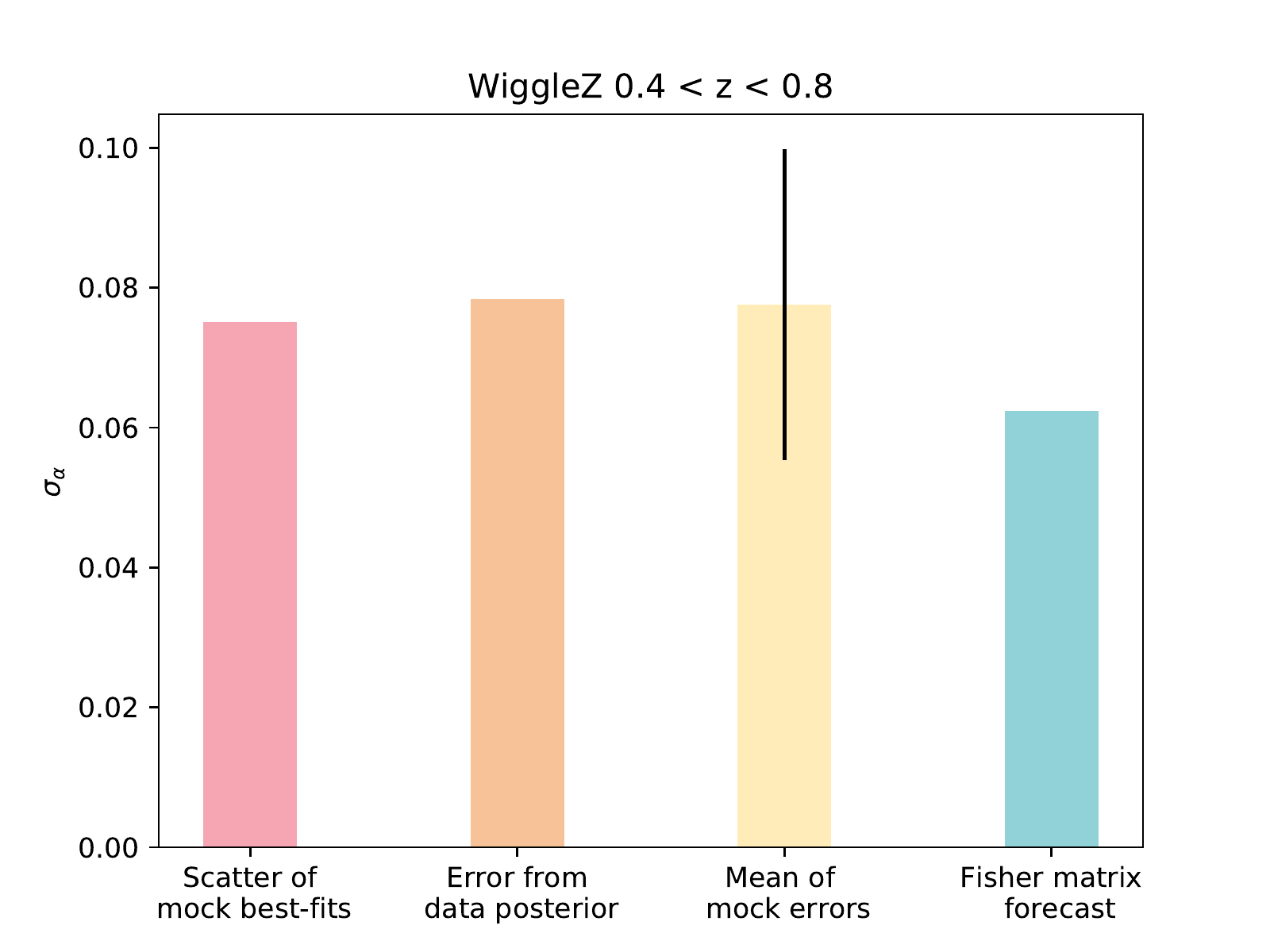}

          \includegraphics[scale =0.5]{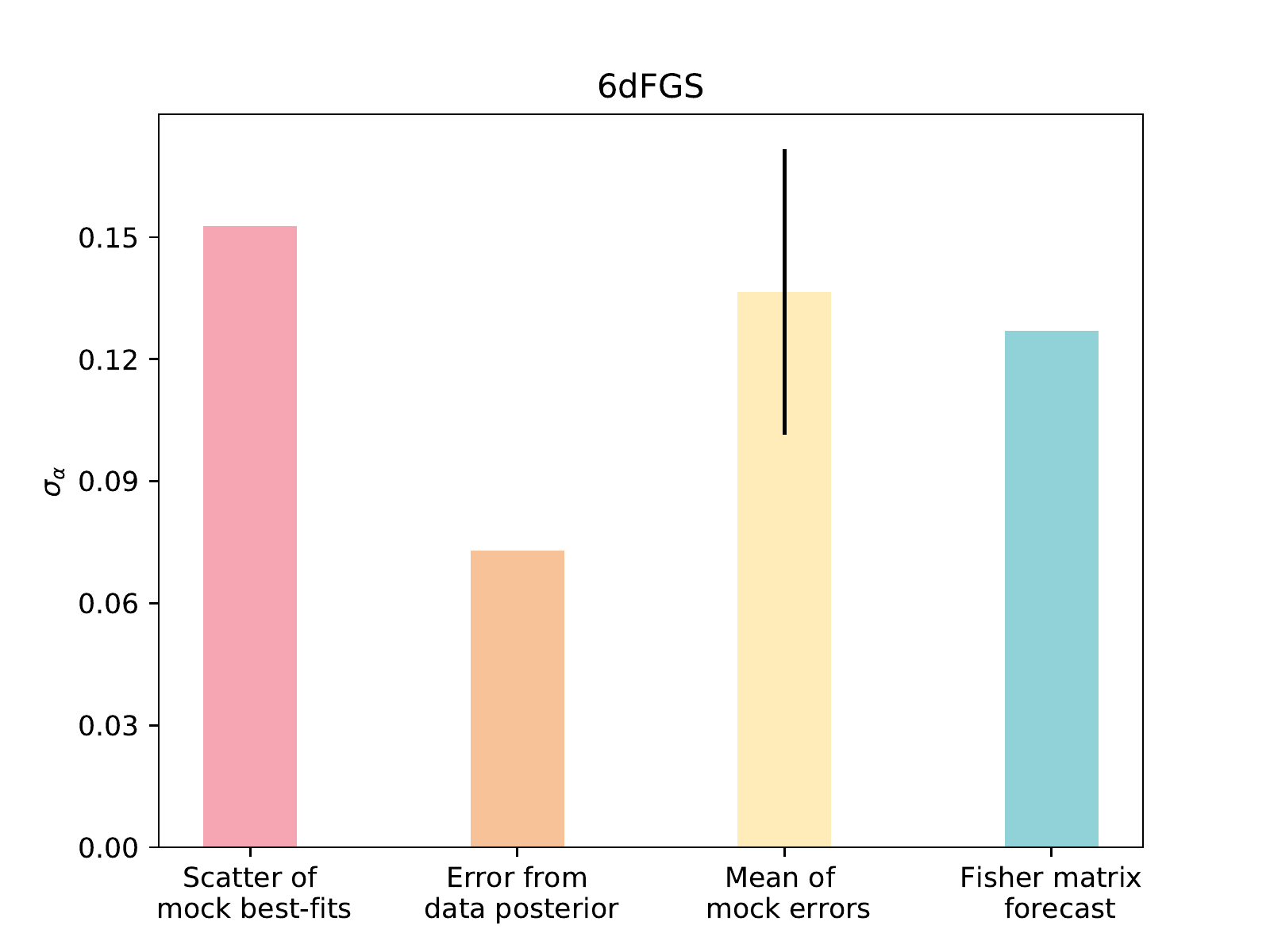}
    \includegraphics[scale =0.5]{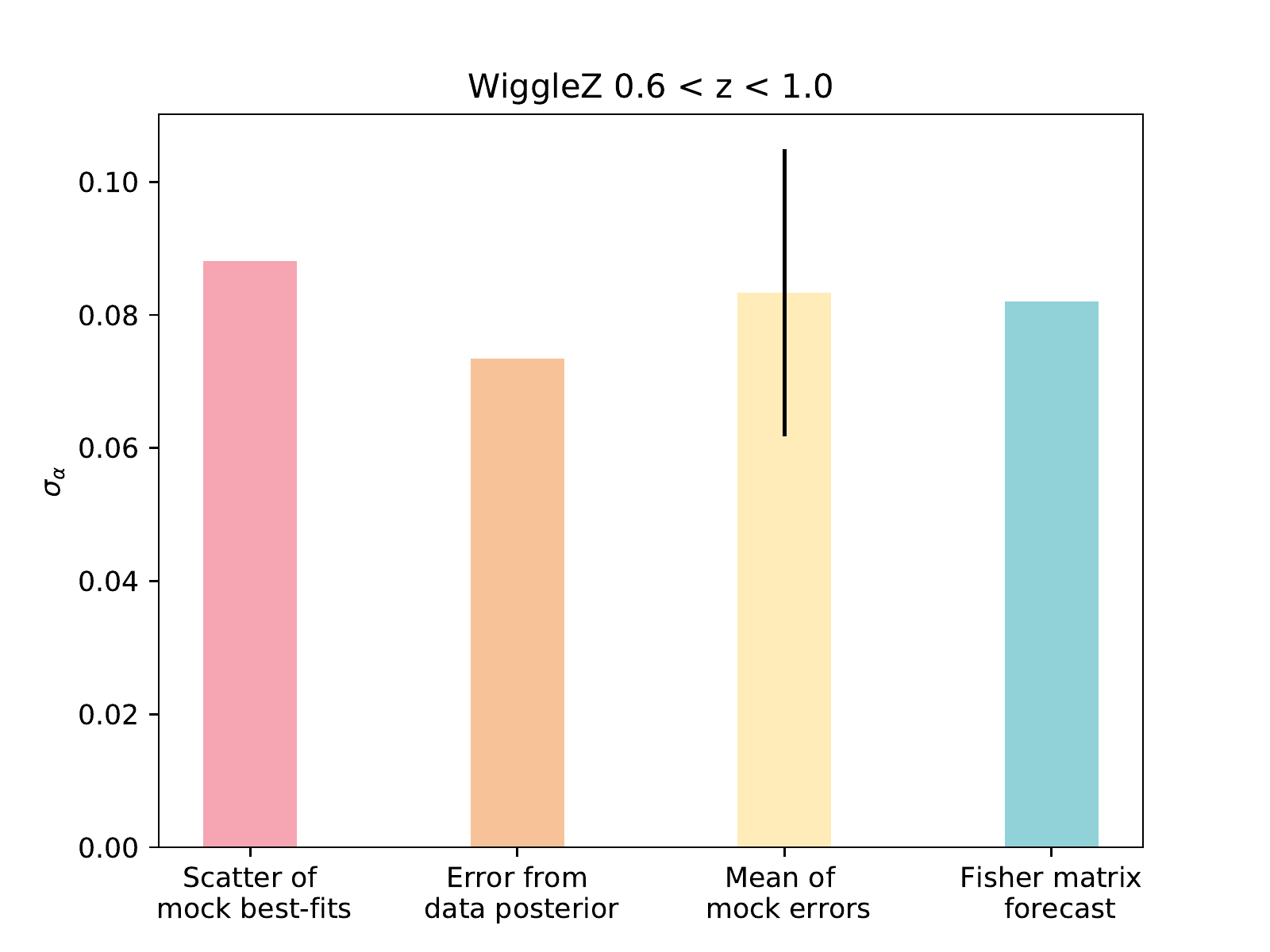}

 %\end{multicols}
    \end{minipage}
  
             \caption{Overall comparison of the different BAO measurement errors for each galaxy survey.  For each panel, we show from left to right: the dispersion of the maximum-likelihood values for $\alpha$ across the mocks (Case 2a), the width of the $68\%$ confidence region $\sigma_\alpha$ from the posterior of the data (Case 1); the average of $\sigma_\alpha$ across the mock posteriors (Case 2b) and the Fisher matrix estimate (Case 3).  The vertical black line in the 3rd column shows the dispersion of the $\sigma_\alpha$ values across the individual mocks.}
     \label{fig:overallcomp}
\end{figure*}

In this paper we have applied a series of different approaches to determine the error in BAO distance measurements from each of the main existing galaxy redshift surveys: 6dFGS, WiggleZ, BOSS and eBOSS.  We adopt three different approaches, described in Section \ref{subsec:error}, comparing the width of the data posterior, the standard deviation across the mock realizations, and a Fisher matrix forecast.  In Figure \ref{fig:overallcomp} we compare these different estimates of the error for the different surveys considered. 

The first column  of the plots corresponds to the error estimated from the scatter of the best-fitting values across the mock realizations (Case 2a) and the second column shows the error derived from the width of the $68\%$ confidence region of the data posterior (Case 1). The error in the third column is computed  averaging  all $68\%$ confidence regions for the individual posteriors from each mock (Case 2b).  The final column corresponds to the error estimated from the Fisher matrix assuming a Gaussian likelihood (Case 3). In general we find  agreement between errors for Cases 2a, 2b, 3: note that if the posteriors are non-Gaussian, then Cases 2a and 2b may produce different results. 

The Fisher matrix error defines a lower limit in the uncertainty of a given parameter, in the Gaussian approximation: the errors obtained from the posterior of the data for 6dFGS, CMASS NGC and WiggleZ ($0.6-1.0$) are lower than this limit. In particular, the posterior of the 6dFGS data produces an error significantly lower than the other estimates.  This is due to the noise component of the error: as we discussed in Sec. \ref{sec:error}, the error of an ML estimator is determined by averaging over all the possible realizations. We refer to this noise as \textit{error in the error}. A crude estimation of this \textit{error in the error} component is given by  the standard deviation of the distribution of the widths of the $68\%$ confidence regions of the individual realizations (black line on the third column of Figure \ref{fig:overallcomp}), which measures the dispersion of the Case 2b errors (third column).

\begin{figure}
    \centering
           \includegraphics[width =\columnwidth]{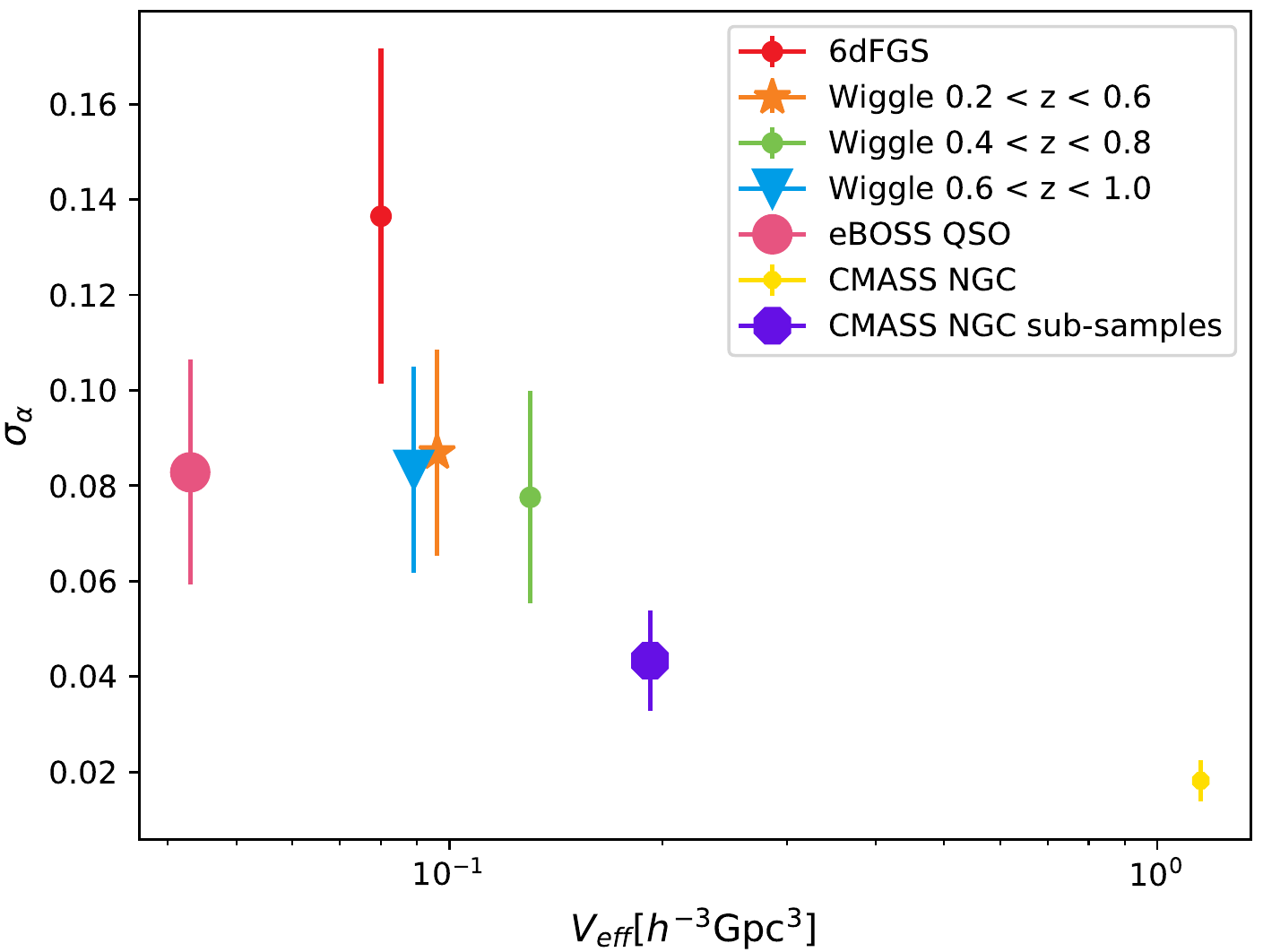}
    \caption{Comparison of the results for all the analyses: 6dFGS ($z \sim$ 0.097), the WiggleZ survey ($z \sim 0.44$, $z\sim 0.60$, $z \sim 0.73$), BOSS CMASS NGC ($z \sim 0.51$), BOSS CMASS NGC sub-samples (average of the 6 sub-sample analyses) and eBOSS QSO ($z \sim 1.52$).  We show the error in the BAO distortion scale $\alpha$ obtained for each survey, as a function of the survey effective volume (x-axis).  The error bars indicate the \textit{error in the error}, i.e. the standard deviation of the distribution of widths inferred from the mock posteriors. %\cab{[Suggest adding units $h^{-3}$ Gpc$^3$ to $x-$axis, change 6DdGFS $\rightarrow$ 6dFGS and Wiggle $\rightarrow$ WiggleZ in legend, can't see the error on the CMASS point -- perhaps move to logarithmic $y$-axis as well?]}
    }
    \label{fig:lastplot}
\end{figure}

Figure \ref{fig:lastplot} compares the errors in the BAO distortion scales obtained from the different surveys, together with the \textit{error in the error} derived from the mock catalogues, as a function of the effective volume of the surveys.  We plot results for  6dFGS ($z \sim$ 0.097), the WiggleZ survey ($z \sim 0.44$, $z\sim 0.60$, $z \sim 0.73$), BOSS CMASS NGC ($z \sim 0.51$), BOSS CMASS NGC sub-samples (corresponding to the average of the results obtained for the 6 sub-samples) and eBOSS QSO ($z \sim 1.52$).  We find that the \textit{error in the error}, driven by sample variance, is more important for low-volume surveys; this is evident from the comparison of the sub-samples and full sample analyses of the CMASS NGC catalogue. The effective volume of the sub-samples is reduced by  a factor of $\sim 6 $ with respect to the full sample, and consequently we observe a significant increase in the \textit{error in the error} from $\sigma_{\alpha}$ $=$ 0.018 $\pm$ 0.004 in the full sample to  $\sigma_{\alpha}$ (sub) $=$ 0.043 $\pm$ 0.016 in the sub-samples.  For the fractional error we get $0.245$ in the sub-samples and $0.2399$ for the full sample, as expected as both $\sigma_{\alpha}$ and $\sigma_{\sigma_{\alpha}}$ scale with the square-root of the sample size. 
%$\sqrt{N}$, with $N$ size of the sample}.
%\cab{[Do we want to say anything about the fractional error in the error?]}

Future surveys such as DESI and Euclid, mapping larger volumes, are expected to limit the sample noise component in the posterior.  However, for low-volume surveys or sub-samples significantly affected by sample variance, multiple definitions of the baryon acoustic peak error may need to be quoted to fully characterize the data and enable accurate comparisons with the cosmological model. 
%\cab{[Is my suggested phrasing O.K. for the final paragraph?]}

% \begin{itemize}
%     \item The error in an ML estimator is determined as an average over all possible
% realisations.
%     \item There is an “error in the error” if BAO errors are determined from the data
% $P(\alpha)$ alone.
%     \item This error in the error is more significant for low-volume surveys (6dFGS)
% than high-volume surveys (BOSS).
%     \item Compare variation in errors for each survey.
% \end{itemize}

\section*{Acknowledgements}
We thank Dr. Florian Beutler,  Prof. Will Percival, Dr. Edward Taylor and Dr. Wael Farah for thoughtful comments and useful discussions.  This research was funded by the Australian Government through Australian Research Council Discovery Project DP160102705 and the Australian Research Council Centre for All-sky Astrophysics (CAASTRO) through project number CE110001020.

\def\jnl@style{\it}
%commente par Seb
\def\aaref@jnl#1{{\jnl@style#1}}
%ref remplace par aaref pour eviter conflit...

\def\aaref@jnl#1{{\jnl@style#1}}

\def\aj{\aaref@jnl{AJ}}                   % Astronomical Journal
\def\araa{\aaref@jnl{ARA\&A}}             % Annual Review of Astron and Astrophys
\def\apj{\aaref@jnl{ApJ}}                 % Astrophysical Journal
\def\apjl{\aaref@jnl{ApJ}}                % Astrophysical Journal, Letters
\def\apjs{\aaref@jnl{ApJS}}               % Astrophysical Journal, Supplement
\def\ao{\aaref@jnl{Appl.~Opt.}}           % Applied Optics
\def\apss{\aaref@jnl{Ap\&SS}}             % Astrophysics and Space Science
\def\aap{\aaref@jnl{A\&A}}                % Astronomy and Astrophysics
\def\aapr{\aaref@jnl{A\&A~Rev.}}          % Astronomy and Astrophysics Reviews
\def\aaps{\aaref@jnl{A\&AS}}              % Astronomy and Astrophysics, Supplement
\def\azh{\aaref@jnl{AZh}}                 % Astronomicheskii Zhurnal
\def\baas{\aaref@jnl{BAAS}}               % Bulletin of the AAS
\def\jrasc{\aaref@jnl{JRASC}}             % Journal of the RAS of Canada
\def\memras{\aaref@jnl{MmRAS}}            % Memoirs of the RAS
\def\mnras{\aaref@jnl{MNRAS}}             % Monthly Notices of the RAS
\def\pra{\aaref@jnl{Phys.~Rev.~A}}        % Physical Review A: General Physics
\def\prb{\aaref@jnl{Phys.~Rev.~B}}        % Physical Review B: Solid State
\def\prc{\aaref@jnl{Phys.~Rev.~C}}        % Physical Review C
\def\prd{\aaref@jnl{Phys.~Rev.~D}}        % Physical Review D
\def\pre{\aaref@jnl{Phys.~Rev.~E}}        % Physical Review E
\def\prl{\aaref@jnl{Phys.~Rev.~Lett.}}    % Physical Review Letters
\def\pasp{\aaref@jnl{PASP}}               % Publications of the ASP
\def\pasj{\aaref@jnl{PASJ}}               % Publications of the ASJ
\def\qjras{\aaref@jnl{QJRAS}}             % Quarterly Journal of the RAS
\def\skytel{\aaref@jnl{S\&T}}             % Sky and Telescope
\def\solphys{\aaref@jnl{Sol.~Phys.}}      % Solar Physics
\def\sovast{\aaref@jnl{Soviet~Ast.}}      % Soviet Astronomy
\def\ssr{\aaref@jnl{Space~Sci.~Rev.}}     % Space Science Reviews
\def\zap{\aaref@jnl{ZAp}}                 % Zeitschrift fuer Astrophysik
\def\nat{\aaref@jnl{Nature}}              % Nature
\def\iaucirc{\aaref@jnl{IAU~Circ.}}       % IAU Cirulars
\def\aplett{\aaref@jnl{Astrophys.~Lett.}} % Astrophysics Letters
\def\apspr{\aaref@jnl{Astrophys.~Space~Phys.~Res.}}
                % Astrophysics Space Physics Research
\def\bain{\aaref@jnl{Bull.~Astron.~Inst.~Netherlands}} 
                % Bulletin Astronomical Institute of the Netherlands
\def\fcp{\aaref@jnl{Fund.~Cosmic~Phys.}}  % Fundamental Cosmic Physics
\def\gca{\aaref@jnl{Geochim.~Cosmochim.~Acta}}   % Geochimica Cosmochimica Acta
\def\grl{\aaref@jnl{Geophys.~Res.~Lett.}} % Geophysics Research Letters
\def\jcp{\aaref@jnl{J.~Chem.~Phys.}}      % Journal of Chemical Physics
\def\jgr{\aaref@jnl{J.~Geophys.~Res.}}    % Journal of Geophysics Research
\def\jqsrt{\aaref@jnl{J.~Quant.~Spec.~Radiat.~Transf.}}
                % Journal of Quantitiative Spectroscopy and Radiative Transfer
\def\memsai{\aaref@jnl{Mem.~Soc.~Astron.~Italiana}}
                % Mem. Societa Astronomica Italiana
\def\nphysa{\aaref@jnl{Nucl.~Phys.~A}}   % Nuclear Physics A
\def\physrep{\aaref@jnl{Phys.~Rep.}}   % Physics Reports
\def\physscr{\aaref@jnl{Phys.~Scr}}   % Physica Scripta
\def\planss{\aaref@jnl{Planet.~Space~Sci.}}   % Planetary Space Science
\def\procspie{\aaref@jnl{Proc.~SPIE}}   % Proceedings of the SPIE
\def\jcap{\aaref@jnl{J. Cosmology Astropart. Phys.}}
                % Journal of Cosmology and Astroparticle Physics

\let\astap=\aap
\let\apjlett=\apjl
\let\apjsupp=\apjs
\let\applopt=\ao

\newcommand{\mpc}{\, {\rm Mpc}}
\newcommand{\kpc}{\, {\rm kpc}}
\newcommand{\hmpc}{\, h^{-1} \mpc}
\newcommand{\ihmpc}{\, h\, {\rm Mpc}^{-1}}
\newcommand{\ikms}{\, {\rm s\, km}^{-1}}
\newcommand{\kms}{\, {\rm km\, s}^{-1}}
\newcommand{\hkpc}{\, h^{-1} \kpc}
\newcommand{\lya}{Ly$\alpha$\ }
\newcommand{\lyb}{Lyman-$\beta$\ }
\newcommand{\lyaf}{Ly$\alpha$ forest}
\newcommand{\lr}{\lambda_{{\rm rest}}}
\newcommand{\bF}{\bar{F}}
\newcommand{\bS}{\bar{S}}
\newcommand{\bC}{\bar{C}}
\newcommand{\bB}{\bar{B}}
\newcommand{\vdF}{{\mathbf \delta_F}}
\newcommand{\vdS}{{\mathbf \delta_S}}
\newcommand{\vdf}{{\mathbf \delta_f}}
\newcommand{\vdn}{{\mathbf \delta_n}}
\newcommand{\vdC}{{\mathbf \delta_C}}
\newcommand{\vdX}{{\mathbf \delta_X}}
\newcommand{\xrei}{x_{rei}}
\newcommand{\lrmin}{\lambda_{{\rm rest, min}}}
\newcommand{\lrmax}{\lambda_{{\rm rest, max}}}
\newcommand{\lmin}{\lambda_{{\rm min}}}
\newcommand{\lmax}{\lambda_{{\rm max}}}
\newcommand{\hi}{\mbox{H\,{\scriptsize I}\ }}
\newcommand{\heii}{\mbox{He\,{\scriptsize II}\ }}
\newcommand{\vp}{\mathbf{p}}
\newcommand{\vq}{\mathbf{q}}
\newcommand{\vxperp}{\mathbf{x_\perp}}
\newcommand{\vkperp}{\mathbf{k_\perp}}
\newcommand{\vrperp}{\mathbf{r_\perp}}
\newcommand{\vx}{\mathbf{x}}
\newcommand{\vy}{\mathbf{y}}
\newcommand{\vk}{\mathbf{k}}
\newcommand{\vR}{\mathbf{r}}
\newcommand{\tdtwo}{\tilde{b}_{\delta^2}}
\newcommand{\tstwo}{\tilde{b}_{s^2}}
\newcommand{\tbthree}{\tilde{b}_3}
\newcommand{\tadtwo}{\tilde{a}_{\delta^2}}
\newcommand{\tastwo}{\tilde{a}_{s^2}}
\newcommand{\tabthree}{\tilde{a}_3}
\newcommand{\vnabla}{\mathbf{\nabla}}
\newcommand{\tpsi}{\tilde{\psi}}
\newcommand{\vv}{\mathbf{v}}
\newcommand{\fnl}{{f_{\rm NL}}}
\newcommand{\tfnl}{{\tilde{f}_{\rm NL}}}
\newcommand{\gnl}{g_{\rm NL}}
\newcommand{\orderfour}{\mathcal{O}\left(\delta_1^4\right)}
\newcommand{\SDSSPF}{\cite{2006ApJS..163...80M}}
\newcommand{\PF}{$P_F^{\rm 1D}(k_\parallel,z)$}
\newcommand\ionalt[2]{#1$\;${\scriptsize \uppercase\expandafter{\romannumeral #2}}}%  
\newcommand{\vxone}{\mathbf{x_1}}
\newcommand{\vxtwo}{\mathbf{x_2}}
\newcommand{\vRot}{\mathbf{r_{12}}}
\newcommand{\cm}{\, {\rm cm}}

\bibliography{bibbao}

\begin{thebibliography}{}
\makeatletter
\relax
\def\mn@urlcharsother{\let\do\@makeother \do\$\do\&\do\#\do\^\do\_\do\%\do\~}
\def\mn@doi{\begingroup\mn@urlcharsother \@ifnextchar [ {\mn@doi@}
  {\mn@doi@[]}}
\def\mn@doi@[#1]#2{\def\@tempa{#1}\ifx\@tempa\@empty \href
  {http://dx.doi.org/#2} {doi:#2}\else \href {http://dx.doi.org/#2} {#1}\fi
  \endgroup}
\def\mn@eprint#1#2{\mn@eprint@#1:#2::\@nil}
\def\mn@eprint@arXiv#1{\href {http://arxiv.org/abs/#1} {{\tt arXiv:#1}}}
\def\mn@eprint@dblp#1{\href {http://dblp.uni-trier.de/rec/bibtex/#1.xml}
  {dblp:#1}}
\def\mn@eprint@#1:#2:#3:#4\@nil{\def\@tempa {#1}\def\@tempb {#2}\def\@tempc
  {#3}\ifx \@tempc \@empty \let \@tempc \@tempb \let \@tempb \@tempa \fi \ifx
  \@tempb \@empty \def\@tempb {arXiv}\fi \@ifundefined
  {mn@eprint@\@tempb}{\@tempb:\@tempc}{\expandafter \expandafter \csname
  mn@eprint@\@tempb\endcsname \expandafter{\@tempc}}}

\bibitem[\protect\citeauthoryear{{Alam} et~al.,}{{Alam}
  et~al.}{2015}]{2015ApJS..219...12A}
{Alam} S.,  et~al., 2015, \mn@doi [\apjs] {10.1088/0067-0049/219/1/12}, \href
  {http://adsabs.harvard.edu/abs/2015ApJS..219...12A} {219, 12}

\bibitem[\protect\citeauthoryear{{Anderson} et~al.,}{{Anderson}
  et~al.}{2014}]{2014MNRAS.441...24A}
{Anderson} L.,  et~al., 2014, \mn@doi [\mnras] {10.1093/mnras/stu523}, \href
  {http://adsabs.harvard.edu/abs/2014MNRAS.441...24A} {441, 24}

\bibitem[\protect\citeauthoryear{{Ata} et~al.,}{{Ata}
  et~al.}{2018}]{2018MNRAS.473.4773A}
{Ata} M.,  et~al., 2018, \mn@doi [\mnras] {10.1093/mnras/stx2630}, \href
  {http://adsabs.harvard.edu/abs/2018MNRAS.473.4773A} {473, 4773}

\bibitem[\protect\citeauthoryear{{Beutler} et~al.,}{{Beutler}
  et~al.}{2011}]{2011MNRAS.416.3017B}
{Beutler} F.,  et~al., 2011, \mn@doi [\mnras]
  {10.1111/j.1365-2966.2011.19250.x}, \href
  {http://adsabs.harvard.edu/abs/2011MNRAS.416.3017B} {416, 3017}

\bibitem[\protect\citeauthoryear{{Blake} et~al.,}{{Blake}
  et~al.}{2011}]{2011MNRAS.415.2892B}
{Blake} C.,  et~al., 2011, \mn@doi [\mnras] {10.1111/j.1365-2966.2011.19077.x},
  \href {http://adsabs.harvard.edu/abs/2011MNRAS.415.2892B} {415, 2892}

\bibitem[\protect\citeauthoryear{{Carlson} \& {White}}{{Carlson} \&
  {White}}{2010}]{2010ApJS..190..311C}
{Carlson} J.,  {White} M.,  2010, \mn@doi [\apjs]
  {10.1088/0067-0049/190/2/311}, \href
  {https://ui.adsabs.harvard.edu/abs/2010ApJS..190..311C} {190, 311}

\bibitem[\protect\citeauthoryear{{Carter}, {Beutler}, {Percival}, {Blake},
  {Koda}  \& {Ross}}{{Carter} et~al.}{2018}]{2018MNRAS.481.2371C}
{Carter} P.,  {Beutler} F.,  {Percival} W.~J.,  {Blake} C.,  {Koda} J.,
  {Ross} A.~J.,  2018, \mn@doi [\mnras] {10.1093/mnras/sty2405}, \href
  {https://ui.adsabs.harvard.edu/abs/2018MNRAS.481.2371C} {481, 2371}

\bibitem[\protect\citeauthoryear{{Cole} et~al.,}{{Cole}
  et~al.}{2005}]{2005MNRAS.362..505C}
{Cole} S.,  et~al., 2005, \mn@doi [\mnras] {10.1111/j.1365-2966.2005.09318.x},
  \href {http://adsabs.harvard.edu/abs/2005MNRAS.362..505C} {362, 505}

\bibitem[\protect\citeauthoryear{{Dawson} et~al.,}{{Dawson}
  et~al.}{2016}]{2016AJ....151...44D}
{Dawson} K.~S.,  et~al., 2016, \mn@doi [\aj] {10.3847/0004-6256/151/2/44},
  \href {https://ui.adsabs.harvard.edu/abs/2016AJ....151...44D} {151, 44}

\bibitem[\protect\citeauthoryear{{Drinkwater} et~al.,}{{Drinkwater}
  et~al.}{2010}]{2010MNRAS.401.1429D}
{Drinkwater} M.~J.,  et~al., 2010, \mn@doi [\mnras]
  {10.1111/j.1365-2966.2009.15754.x}, \href
  {https://ui.adsabs.harvard.edu/abs/2010MNRAS.401.1429D} {401, 1429}

\bibitem[\protect\citeauthoryear{{Eisenstein} \& {Hu}}{{Eisenstein} \&
  {Hu}}{1998}]{1998ApJ...496..605E}
{Eisenstein} D.~J.,  {Hu} W.,  1998, \mn@doi [\apj] {10.1086/305424}, \href
  {http://adsabs.harvard.edu/abs/1998ApJ...496..605E} {496, 605}

\bibitem[\protect\citeauthoryear{{Eisenstein} et~al.,}{{Eisenstein}
  et~al.}{2005}]{2005ApJ...633..560E}
{Eisenstein} D.~J.,  et~al., 2005, \mn@doi [\apj] {10.1086/466512}, \href
  {http://adsabs.harvard.edu/abs/2005ApJ...633..560E} {633, 560}

\bibitem[\protect\citeauthoryear{{Feldman}, {Kaiser}  \& {Peacock}}{{Feldman}
  et~al.}{1994}]{1994ApJ...426...23F}
{Feldman} H.~A.,  {Kaiser} N.,   {Peacock} J.~A.,  1994, \mn@doi [\apj]
  {10.1086/174036}, \href {http://adsabs.harvard.edu/abs/1994ApJ...426...23F}
  {426, 23}

\bibitem[\protect\citeauthoryear{{Font-Ribera}, {McDonald}, {Mostek}, {Reid},
  {Seo}  \& {Slosar}}{{Font-Ribera} et~al.}{2014}]{2014JCAP...05..023F}
{Font-Ribera} A.,  {McDonald} P.,  {Mostek} N.,  {Reid} B.~A.,  {Seo} H.-J.,
  {Slosar} A.,  2014, \mn@doi [Journal of Cosmology and Astro-Particle Physics]
  {10.1088/1475-7516/2014/05/023}, \href
  {https://ui.adsabs.harvard.edu/abs/2014JCAP...05..023F} {2014, 023}

\bibitem[\protect\citeauthoryear{Gelman \& Rubin}{Gelman \&
  Rubin}{1992}]{Gelman:1992zz}
Gelman A.,  Rubin D.~B.,  1992, \mn@doi [Statist. Sci.]
  {10.1214/ss/1177011136}, 7, 457

\bibitem[\protect\citeauthoryear{{Heavens }}{{Heavens }}{2011}]{alahheavlect}
{Heavens } A.,  2011, Lectures and workshops at Penn State, June 2011

\bibitem[\protect\citeauthoryear{{Hogg} \& {Foreman-Mackey}}{{Hogg} \&
  {Foreman-Mackey}}{2018}]{2018ApJS..236...11H}
{Hogg} D.~W.,  {Foreman-Mackey} D.,  2018, \mn@doi [\apjs]
  {10.3847/1538-4365/aab76e}, \href
  {https://ui.adsabs.harvard.edu/abs/2018ApJS..236...11H} {236, 11}

\bibitem[\protect\citeauthoryear{{Hogg}, {Bovy}  \& {Lang}}{{Hogg}
  et~al.}{2010}]{2010arXiv1008.4686H}
{Hogg} D.~W.,  {Bovy} J.,   {Lang} D.,  2010, arXiv e-prints, \href
  {https://ui.adsabs.harvard.edu/abs/2010arXiv1008.4686H} {p. arXiv:1008.4686}

\bibitem[\protect\citeauthoryear{{Kazin} et~al.,}{{Kazin}
  et~al.}{2014}]{2014MNRAS.441.3524K}
{Kazin} E.~A.,  et~al., 2014, \mn@doi [\mnras] {10.1093/mnras/stu778}, \href
  {https://ui.adsabs.harvard.edu/abs/2014MNRAS.441.3524K} {441, 3524}

\bibitem[\protect\citeauthoryear{{Koda}, {Blake}, {Beutler}, {Kazin}  \&
  {Marin}}{{Koda} et~al.}{2016}]{2016MNRAS.459.2118K}
{Koda} J.,  {Blake} C.,  {Beutler} F.,  {Kazin} E.,   {Marin} F.,  2016,
  \mn@doi [\mnras] {10.1093/mnras/stw763}, \href
  {https://ui.adsabs.harvard.edu/abs/2016MNRAS.459.2118K} {459, 2118}

\bibitem[\protect\citeauthoryear{{Landy} \& {Szalay}}{{Landy} \&
  {Szalay}}{1993}]{1993ApJ...412...64L}
{Landy} S.~D.,  {Szalay} A.~S.,  1993, \mn@doi [\apj] {10.1086/172900}, \href
  {https://ui.adsabs.harvard.edu/abs/1993ApJ...412...64L} {412, 64}

\bibitem[\protect\citeauthoryear{{Laureijs} et~al.,}{{Laureijs}
  et~al.}{2011}]{2011arXiv1110.3193L}
{Laureijs} R.,  et~al., 2011, arXiv e-prints, \href
  {https://ui.adsabs.harvard.edu/abs/2011arXiv1110.3193L} {p. arXiv:1110.3193}

\bibitem[\protect\citeauthoryear{{Levi} et~al.,}{{Levi}
  et~al.}{2019}]{2019arXiv190710688L}
{Levi} M.~E.,  et~al., 2019, arXiv e-prints, \href
  {https://ui.adsabs.harvard.edu/abs/2019arXiv190710688L} {p. arXiv:1907.10688}

\bibitem[\protect\citeauthoryear{{Planck Collaboration} et~al.,}{{Planck
  Collaboration} et~al.}{2016}]{2016A&A...594A..13P}
{Planck Collaboration} et~al., 2016, \mn@doi [\aap]
  {10.1051/0004-6361/201525830}, \href
  {http://adsabs.harvard.edu/abs/2016A%26A...594A..13P} {594, A13}

\bibitem[\protect\citeauthoryear{{Ross} et~al.,}{{Ross}
  et~al.}{2017}]{2017MNRAS.464.1168R}
{Ross} A.~J.,  et~al., 2017, \mn@doi [\mnras] {10.1093/mnras/stw2372}, \href
  {https://ui.adsabs.harvard.edu/abs/2017MNRAS.464.1168R} {464, 1168}

\bibitem[\protect\citeauthoryear{{Seo} \& {Eisenstein}}{{Seo} \&
  {Eisenstein}}{2007}]{2007ApJ...665...14S}
{Seo} H.-J.,  {Eisenstein} D.~J.,  2007, \mn@doi [\apj] {10.1086/519549}, \href
  {https://ui.adsabs.harvard.edu/abs/2007ApJ...665...14S} {665, 14}

\bibitem[\protect\citeauthoryear{{Slosar} et~al.,}{{Slosar}
  et~al.}{2013}]{2013JCAP...04..026S}
{Slosar} A.,  et~al., 2013, \mn@doi [Journal of Cosmology and Astro-Particle
  Physics] {10.1088/1475-7516/2013/04/026}, \href
  {https://ui.adsabs.harvard.edu/abs/2013JCAP...04..026S} {2013, 026}

\bibitem[\protect\citeauthoryear{{Tassev}, {Zaldarriaga}  \&
  {Eisenstein}}{{Tassev} et~al.}{2013}]{2013JCAP...06..036T}
{Tassev} S.,  {Zaldarriaga} M.,   {Eisenstein} D.~J.,  2013, \mn@doi [Journal
  of Cosmology and Astro-Particle Physics] {10.1088/1475-7516/2013/06/036},
  \href {https://ui.adsabs.harvard.edu/abs/2013JCAP...06..036T} {2013, 036}

\bibitem[\protect\citeauthoryear{{Tinker} et~al.,}{{Tinker}
  et~al.}{2012}]{2012ApJ...745...16T}
{Tinker} J.~L.,  et~al., 2012, \mn@doi [\apj] {10.1088/0004-637X/745/1/16},
  \href {https://ui.adsabs.harvard.edu/abs/2012ApJ...745...16T} {745, 16}

\bibitem[\protect\citeauthoryear{{Trotta}}{{Trotta}}{2017}]{2017arXiv170101467T}
{Trotta} R.,  2017, arXiv e-prints, \href
  {https://ui.adsabs.harvard.edu/abs/2017arXiv170101467T} {p. arXiv:1701.01467}

\bibitem[\protect\citeauthoryear{{Weinberg}, {Mortonson}, {Eisenstein},
  {Hirata}, {Riess}  \& {Rozo}}{{Weinberg} et~al.}{2013}]{2013PhR...530...87W}
{Weinberg} D.~H.,  {Mortonson} M.~J.,  {Eisenstein} D.~J.,  {Hirata} C.,
  {Riess} A.~G.,   {Rozo} E.,  2013, \mn@doi [\physrep]
  {10.1016/j.physrep.2013.05.001}, \href
  {http://adsabs.harvard.edu/abs/2013PhR...530...87W} {530, 87}

\bibitem[\protect\citeauthoryear{{White} et~al.,}{{White}
  et~al.}{2011}]{2011ApJ...728..126W}
{White} M.,  et~al., 2011, \mn@doi [\apj] {10.1088/0004-637X/728/2/126}, \href
  {https://ui.adsabs.harvard.edu/abs/2011ApJ...728..126W} {728, 126}

\bibitem[\protect\citeauthoryear{{White}, {Tinker}  \& {McBride}}{{White}
  et~al.}{2014}]{2014MNRAS.437.2594W}
{White} M.,  {Tinker} J.~L.,   {McBride} C.~K.,  2014, \mn@doi [\mnras]
  {10.1093/mnras/stt2071}, \href
  {http://adsabs.harvard.edu/abs/2014MNRAS.437.2594W} {437, 2594}

\makeatother
\end{thebibliography}

%%%%%%%%%%%%%%%%%%%%%%%%%%%%%%%%%%%%%%%%%%%%%%%%%%

%%%%%%%%%%%%%%%%% APPENDICES %%%%%%%%%%%%%%%%%%%%%

\appendix

% \section{Some extra material}

% If you want to present additional material which would interrupt the flow of the main paper,
% it can be placed in an Appendix which appears after the list of references.

%%%%%%%%%%%%%%%%%%%%%%%%%%%%%%%%%%%%%%%%%%%%%%%%%%

% Don't change these lines
\bsp	% typesetting comment
%\label{lastpage}
\end{document}